\documentclass[11pt]{article}

\usepackage{amsmath}  
\usepackage{amsfonts,dsfont}
\usepackage{amssymb}
\usepackage{amsthm}
\usepackage{mathtools}
\usepackage{thmtools}
\usepackage{bm} 

\usepackage{graphicx}
\usepackage[hidelinks]{hyperref}
\usepackage[english]{babel}

\newlength{\bredde}
\def\slash#1{\settowidth{\bredde}{$#1$}\ifmmode\,\raisebox{.15ex}{/}
\hspace*{-\bredde} #1\else$\,\raisebox{.15ex}{/}\hspace*{-\bredde} #1$\fi}
\newcommand{\diag}{{\rm diag\,}}

\newcommand{\tr}{{\rm tr\,}}
\newcommand{\Tr}{{\rm Tr\,}}

\newcommand{\eins}{\leavevmode\hbox{\small1\kern-3.8pt\normalsize1}}
\newcommand{\Res}{{\rm Res\,}}
\newcommand{\be}{\begin{equation}}
\newcommand{\ee}{\end{equation}}
\newcommand{\bee}{\begin{eqnarray}}
\newcommand{\eee}{\end{eqnarray}}

\newcommand{\sect}[1]{\setcounter{equation}{0}\section{#1}}

\begin{document}

\title{Universal microscopic spectrum of the unquenched QCD Dirac operator at finite temperature}
\author{{\sc G. Akemann$^{1,2}$, T.R. W\"urfel$^{1,3}$}\\~\\
$^1$Faculty of Physics, Bielefeld University,
Postfach 100131, D-33501 Bielefeld, Germany\\
$^2$Mathematical Sciences Research Institute,  17 Gauss Way,
Berkeley, CA 94720-5070, USA\\
$^3$Department of Mathematics, King's College London, London WC2R 2LS, UK}
\date{}
\maketitle
\begin{abstract}
In the $\varepsilon$-regime of chiral perturbation theory the spectral correlations of the Euclidean QCD Dirac operator close to the origin can be computed using random matrix theory. To incorporate the effect of temperature, a random matrix ensemble has been proposed, where a constant, deterministic matrix is added to the Dirac operator. 
Its eigenvalue correlation functions can be written as the determinant of a kernel that depends on temperature. 
Due to recent progress in this specific class of random matrix ensembles, featuring a deterministic, 
additive shift, we can determine the limiting kernel and correlation functions in this class, which is the class of polynomial ensembles.
We prove the equivalence between this new determinantal representation 
of the microscopic eigenvalue correlation functions 
and existing results in terms of determinants of different sizes, for an arbitrary number of quark flavours, with and without temperature, and extend them to non-zero topology.  
These results all agree and are thus universal when measured in units of the temperature dependent chiral condensate, as long as we stay below the chiral phase transition.

\end{abstract}

\sect{Introduction}\label{sec:intro}

The application of random matrices to the low energy spectrum of the Euclidean QCD Dirac operator spectrum has been very successful, starting from the seminal paper of Shuryak and Verbaarschot \cite{SV}. It is surprising how many details of the underlying field theory have been understood analytically within this description, which is sometimes called the microscopic limit of QCD. 
The correspondence between different gauge groups and their representations, and random matrix ensembles was established \cite{Jac3fold}, based on global anti-unitary symmetries of the underlying field theory in the continuum.
This classification was recently revisited on the lattice, in various dimensions \cite{KVZ, KVZ2}. It was understood how to include a fixed number of quark flavours $N_f$ 
for all spectral correlation functions
\cite{DN,WGW}, and the universality beyond using random matrices with a Gaussian distribution was established \cite{DN, ADMN}. The corrections from a finite lattice spacing were included into a random matrix ensemble \cite{DSV,ADSV, Mario}. Similarly, a quark chemical potential $\mu$ was included in the ensemble \cite{Misha,James}, and all complex eigenvalue correlation functions of the Dirac operator were be computed \cite{James}, despite the sign problem. 

Perhaps most importantly, it was understood which limit has to be taken starting from field theory, in order to arrive at a random matrix description. The limit is given by the $\varepsilon$-regime of chiral perturbation theory ($\varepsilon\chi$PT), that was defined before by Gasser and Leutwyler \cite{GL} as a regime that is analytically tractable. The equivalence between random matrix statistics and leading order $\varepsilon\chi$PT was established for the microscopic correlation functions \cite{DOTV,DOTV2,TV,BA}, including $\mu\neq0$ \cite{BA}. The mapping to Wilson $\varepsilon\chi$PT at finite lattice spacing was achieved for the microscopic density in \cite{DSV,ADSV}. 

At  next-to-leading order in $\varepsilon\chi$PT the low energy constants (LEC) merely have to be renormalised by finite volume corrections \cite{DDF,DDF2,DDF3}, keeping the equivalence to random matrices intact.
Deviations from random matrix statistics were found starting from next-to-next-to leading order in \cite{LHW}.  Although $\varepsilon\chi$PT is an unphysical limit, its predictions have been very useful for instance in determining the LEC in chiral perturbation theory from lattice data, such as the chiral condensate \cite{Sigma}, or the Pion decay constant \cite{FPi}. 
After the development of Fermion discretisations that preserve topology, a comparison to the detailed, topology dependent predictions became possible \cite{urs}.
Furthermore, a systematic expansion has been established that allows to interpolate from the $\varepsilon$- to the $p$-regime  for the microscopic density \cite{PoulHide}. For more details and references beyond this very brief overview we refer to several reviews in the field \cite{VW,Poul2011,GA16}.


What happens when introducing temperature? At low temperature the LEC in the chiral Lagrangian get renormalised. The corrections to the chiral condensate \cite{GL, GLab}, Pion decay constant and Mass \cite{DT} were computed to order $T^4$. In a review by Leutwyler it was concluded, “that $\chi$PT yields reliable results only below a temperature of about $130$ MeV” \cite{HLscholar}. On the random matrix side a temperature dependent ensemble was introduced \cite{JV,WSW}, that led to very important insights. This is the ensemble that we will study in this article. First, it was used as a phenomenological model for the chiral phase transition at $T_c\approx 153\ \text{MeV}$. At zero quark mass a second order phase transition was found and the mean field critical exponents determined \cite{JV,WSW}. The characteristic structure of the phase diagram as a function of $T$ and $\mu$, applicable to QCD with $N_f=2$ light flavours, was predicted in a very influential paper \cite{phase}. The quenched microscopic density at $T_c$ was computed in \cite{Janik}.

Sufficiently below $T_c$, one can expect that the predictions of $\varepsilon\chi$PT remain valid using the renomalised LEC. The quenched microscopic correlation functions of the random matrix Dirac operator were computed \cite{JSVev,GuhrWettig} and found to be independent of temperature, when rescaling the eigenvalues with the temperature dependent chiral condensate $\Xi(T)>0$. More importantly, they agree with the universal predictions for the quenched model at zero temperature \cite{ADMN}. These calculations at $T\neq0$ were extended to 
include a fixed number of quark masses at zero topology in \cite{SWG}. Their results were conjectured to agree with the zero temperature predictions \cite{DN,WGW} as well.
A comparison with lattice simulations below and across $T_c$ was made \cite{LattT,LattT2}, where an analogy to Anderson localisation was suggested \cite{GGO}.


Which questions are interesting to be considered further in a temperature dependent random matrix ensemble? First, as described above, the question was left open whether or not the microscopic Dirac operator eigenvalue correlation functions with and without temperature are in the same universality class below the phase transition, apart from the quenched approximation and numerical checks. We will provide a comprehensive and affirmative 
analytic answer to this question. 
This equivalence includes interesting identities between determinantal representations of different sizes of the same correlation functions. The idea of the proof is based on different representations of expectation values of characteristic polynomials at zero temperature \cite{AV03}, as well as a set of consistency conditions \cite{ADII,Braden} that holds among the finite volume partition functions of $\varepsilon\chi$PT. 

Second, it is still debated within the lattice community, if the axial anomaly $U_A(1)$ is restored at or beyond the chiral phase transition, or if it remains broken. Different groups come to different conclusions \cite{Ed,Hide,Brandt}, see e.g.   \cite{olaf} for a recent discussion and further references. It could be thus useful to have an effective random matrix description at criticality, that allows for a detailed study of the microscopic Dirac operator spectrum, including in particular the role of topology. 
We will provide a detailed solution for all  eigenvalue correlation functions of the random matrix ensemble proposed in \cite{JV,WSW} at finite matrix size $N$,  
for an arbitrary fixed number of quark flavours $N_f$, topology $\nu$ and parameters of the external field that models the effect of temperature. They are given in terms of a determinant of a kernel for which we derive an explicit double contour representation. 
This detailed answer will allow us to address the microscopic correlations at the phase transition in this ensemble at criticality, to which we plan to come back in future work. 

Our results have only become possible due to recent mathematical developments in random matrix theory on so-called polynomial ensembles \cite{DF,FGS,ASW}. 
In the classical ensembles with unitary symmetry the repulsion between random matrix eigenvalues is given by the square of the Vandermonde determinant. The addition of a deterministic matrix, that models the effect of temperature here, partly destroys this structure. However, all eigenvalue correlation functions can still be computed in terms of the determinant of a kernel of bi-orthogonal functions. One set of these remain to be polynomials, explaining the name of these ensembles. 
For certain subclasses of polynomial ensembles compact contour integral representations of the kernel are available \cite{ASW}, that allow for an asymptotic analysis.

The remainder of this article is organised as follows: In Section \ref{sec:model} we recall the random matrix ensemble for the QCD Dirac operator at non-zero temperature, show that it is a polynomial ensemble and construct its corresponding kernel and correlation functions at finite matrix size $N$. The microscopic large-$N$ limit is taken in Section \ref{sec:asymptotic}, where for clarity we first present the saddle point analysis of the quenched kernel in Subsection \ref{subsec:qK}. It tells us when we are in the phase, where chiral symmetry is broken, defining the random matrix order parameter $\Xi>0$. The fully unquenched result for the limiting kernel is given in Subsection \ref{subsec:uqK}. In Section \ref{sec:Universal} we make contact to earlier results, 
and extend these to topology $\nu>0$. In Subsection \ref{subsec:equiv} we establish the equivalence between the known unquenched kernel at zero temperature and our limiting kernel.
Subsection \ref{subsec:equivII} extends this equivalence to previous results for the correlation functions at non-zero temperature and thus establishes the universality of all these different representations below the phase transition. 
Our conclusions and discussion of open problems are offered in Section \ref{sec:conclusio}.

\sect{Ensemble for unquenched QCD at non-zero temperature}\label{sec:model}

In this section we will recall the schematic ensemble \cite{JV,WSW} of random matrices   for QCD at non-zero temperature, with $N_f$ massive quark flavours and fixed topology $\nu\geq0$. We first give the matrix representation and derive the joint density of Dirac operator eigenvalues in Subsection \ref{subsec:jpdf}, which will be the starting point of our analysis. A critical difference regarding the integrable structure of the model at zero and non-zero temperature is pointed out, before in Subsection \ref{subsec:correl} we give the general result for all $k$-point correlation functions at finite matrix size $N$ in terms of the kernel as a double integral, applying results from \cite{ASW}.

\subsection{The matrix representation and its joint eigenvalue distribution}
\label{subsec:jpdf}

The random matrix partition function for low energy QCD with temperature is given by 
\begin{equation}\label{ZNA}
\mathbb{Z}_{N,\nu}^{(N_f)}(A)=\int_{\mathbb{C}^{N\times(N+\nu)}} d[W]\exp[-\Tr WW^\dag]\prod_{f=1}^{N_f}\det[\mathcal{D}(A)+m_f\eins_{2N+\nu}],
\end{equation}
where $W$ is an $N\times(N+\nu)$ matrix with Gaussian distribution, and we integrate with respect to the flat Lebesgue measure $d[W]$ over all independent, complex matrix elements of $W$.
The special ensemble at $N_f=0$ is also known as the \emph{chiral Gaussian unitary ensemble} (chGUE) with an external field \cite{DF}.
The anti-Hermitian Euclidean Dirac operator $\mathcal{D}(A)$ has been Wick-rotated already to have real eigenvalues. Together with its schematic temperature dependence, modelled by a deterministic matrix $A$, the Dirac operator reads 
\begin{eqnarray}\label{diracoptemp}
\mathcal{D}(A) = 
\left( \begin{matrix}
0 & W + A \\
W^\dagger + A^\dag & 0 \\
\end{matrix} \right).
\end{eqnarray}
As we will show below, without loss of generality, we can choose the quadratic part of $A$ to be given by the fixed diagonal matrix $\diag[\sqrt{a_1},\ldots,\sqrt{a_N}]$ and the rest being zero. This means the real and positive parameters $\{a_j\}_{j=1}^N$ are the squared singular values of $A$ (or eigenvalues of $AA^\dag$). We assume them to be pairwise distinct and non-vanishing, $a_j>0$ for all $j=1,\ldots,N$. 
In the initial proposal \cite{JV,WSW} (at $\nu=0$) the matrix $A$ contains the Matsubara frequencies, which come in odd integer multiples of $\pm\pi T$ for Fermions, with $T$ being the temperature. Taking into account only the lowest Matsubara frequency thus reduces the matrix $A$ to $A=\pi T\eins_N$. Here, as in \cite{GuhrWettig}, we will take a more formal point of view,  with \eqref{ZNA} being a schematic model, and allow for arbitrary fixed eigenvalues of $AA^\dag$, as long as we stay in the phase, where chiral symmetry is broken. 

After shifting the random matrix $W\to W^\prime=W+A$ we are led to consider the following matrix integral
\begin{equation}\label{Zshift}
\mathbb{Z}_{N,\nu}^{(N_f)}(A)=\int_{\mathbb{C}^{N\times(N+\nu)}} d[W^\prime] \ 
e^{-\Tr W^\prime W^{\prime\dag}-\Tr AA^\dag+\Tr(W^\prime A^\dag+W^{\prime\dag}A)} \prod_{f=1}^{N_f}m_f^\nu\det[m_f^2\eins_N+W^\prime W^{\prime\dag}],
\end{equation}
cf. \cite{DF} in case $N_f=0$.
When making a singular value decomposition of $W^\prime=UXV^\dag$ (or equivalently diagonalising $W^\prime W^{\prime\dag}$), the matrix $X=(\diag(\sqrt{x_1},\ldots,\sqrt{x_N})\ |\ 0_{N\times\nu})$ is given in terms of the squared singular values $\{x_j\}_{j=1}^N$ and a block matrix of zero entries. 
The eigenvectors no longer decouple in the last two terms in the exponential in \eqref{Zshift}. Therefore, we have to use the following group integral\footnote{The invariance of $W^\prime$ is in fact under the coset $U(N)\times U(N+\nu)/U(1)^N$, but the integration domain can be extended to the full unitary groups.} derived in \cite{JSVgroup,JSVgroup2} for rectangular matrices, also called Berezin-Karpelevich integral formula,
\begin{equation}\label{BKintegral}
\int_{U(N)} d[U] \int_{U(N+\nu)} d[V] \exp[ \Tr (UX V^\dagger A^\dag+ VX^\dag U^\dag A)] = C_{N,\nu} \prod_{j=1}^{N} (a_j x_j)^{-\frac{\nu}{2}}\frac{\det[I_\nu (2\sqrt{a_i x_j})]_{i,j=1}^N}{\Delta_N(\{ a \}) \Delta_N (\{x\})}.
\end{equation}
Here, we integrate over the Haar measures $d[U]$ and $d[V]$ of the corresponding unitary groups, 
\begin{equation}
\label{Vandermonde}
\Delta_N(\{x\})=\prod_{i>j}(x_i-x_j)=\det[x_i^{j-1}]_{i,j=1}^N
\end{equation}
denotes the Vandermonde determinant\footnote{Note that in \cite{ASW} a different convention is used for the Vandermonde determinant, differing by the factor $(-1)^{N(N-1)/2}$.}, and the modified Bessel function of the first kind reads
\begin{equation}\label{Bessel-Idef}
I_\nu(z)=\sum_{n=0}^\infty \frac{(z/2)^{2n+\nu}}{n!\Gamma(n+\nu+1)}.
\end{equation}
The constant $C_{N,\nu}$ is known but does not depend on the set of parameters $\{a_j\}_{j=1}^N$ and is thus not important in the sequel. It is also clear now why we could choose $A$ to be diagonal in the first place: The fixed unitary matrices in a singular value decomposition of a generic $A^\prime=U^\prime A V^{\prime\dag}$ could have been absorbed into $U$ and $V$ by the property of the left and right invariance of the Haar measure, allowing $A$ to become diagonal.

To compute the joint probability distribution function (jpdf) of singular values of $W^\prime$ we need one more ingredient, the corresponding Jacobian. It is well known, and we use this opportunity to define the model at $A=0$, where it also appears:
\begin{eqnarray}\label{ZN0}
\mathbb{Z}_{N,\nu}^{(N_f)}(0)&=&\int_{\mathbb{C}^{N\times(N+\nu)}} d[W]\exp[-\Tr WW^\dag]\prod_{f=1}^{N_f}m_f^\nu\det[m_f^2\eins_{N}+WW^\dag]\\
&\sim& \int_{\mathbb{R}_+^N}d[x]\prod_{f=1}^{N_f}m_f^\nu\left(\prod_{j=1}^N x_j^\nu e^{-x_j}\prod_{f=1}^{N_f}(x_j+m_f^2) \right) \Delta_N(\{x\})^2\nonumber\\
\label{PNf0}
&\equiv& \int_{\mathbb{R}_+^N}d[x]\ \mathcal{P}_{N,\nu}^{N_f}(\{x\})=Z_{N,\nu}^{(N_f)}(\{m\}).
\end{eqnarray}
Here we used $d[x]=\prod_{j=1}^Ndx_j$. In the last line we have defined the (unnormalised) jpdf at zero temperature $\mathcal{P}_{N,\nu}^{N_f}(\{x\})$, and we suppressed the proportionality constant. Note: There are different conventions in the literature, with or without keeping the factor $\prod_{f=1}^{N_f}m_f^\nu$ as part of the jpdf.

Combining this Jacobian for $W$ (or $W^\prime$) with the group integral \eqref{BKintegral}, we arrive at the jpdf $\mathcal{P}_{N,\nu}^{N_f}(\{x\},\{a\})$ for the model with non-zero temperature
\begin{eqnarray}
Z_{N,\nu}^{(N_f)}(\{m\};\{a\})&=&\int_{\mathbb{R}_+^N}d[x]\!
\left(\prod_{f=1}^{N_f}m_f^\nu\prod_{j=1}^N(x_j+m_f^2)\right)\!\Delta_N(\{x\})
\frac{\det\left[\left(\frac{x_j}{a_i}\right)^{\nu/2}e^{-x_j-a_i}I_\nu (2\sqrt{a_i x_j})\right]_{i,j=1}^N}{\Delta_N(\{ a \})}\nonumber\\
\label{PNfA}
&\equiv& \int_{\mathbb{R}_+^N}d[x]\ \mathcal{P}_{N,\nu}^{N_f}(\{x\},\{a\}).
\end{eqnarray}
Above, we have pulled all the $N_f$-independent weight factors into the second determinant in the first line. All $a_j$-dependent factors are displayed explicitly, in order to be able to take the limit $a_j\to0$ to recover the model without temperature above.

\subsection{Correlation functions and kernels at finite-$N$}
\label{subsec:correl}

Both with and without temperature, the jpdf's \eqref{PNf0} and \eqref{PNfA} 
take the form of the product of two determinants. 
They belong to the more general class of bi-orthogonal ensembles \cite{Borodin}
where the two sets of functions inside the determinants are orthonormalised. 
In the mathematical literature this setup is called determinantal point process, meaning that all correlation functions to be defined below can be written as the determinant of the kernel of these bi-orthogonal functions. The name point process comes from the analogy to stochastic processes of $N$ points (the eigenvalues) of this type.

Specifically, at $A=0$ the jpdf  \eqref{PNf0} is called {\it orthogonal polynomial ensemble}, containing two Vandermonde determinants. Here, the kernel can be written in terms of orthogonal polynomials, as we will see. 
The more general jpdf \eqref{PNfA} is called {\it polynomial ensemble}, containing just one Vandermonde determinant, see \cite{Arno} for its definition and general properties. This implies that among the two sets of functions that are bi-orthogonal and constitute the kernel, one set is always given by polynomials.

For both setups, including $A\neq0$,
the $k$-point correlation functions are defined as 
\begin{eqnarray}
\label{RNkdef}
R_{k,A}^{(N_f)}(x_1,\ldots,x_k)&\equiv& \frac{N!}{(N-k)!}\frac{1}{ Z_{N,\nu}^{(N_f)}(\{m\};\{a\})}
\int_{\mathbb{R}_+^{N-k}} dx_{k+1}\cdots dx_N \mathcal{P}_{N,\nu}^{N_f}(\{x\},\{a\})
\nonumber\\
\label{Rk-kernel}
&=&
\det\left[K_{N,A}^{(N_f)}(x_i,x_j)\right]_{i,j=1}^k,
\end{eqnarray}
 and can be expressed in terms of 
a kernel $K_{N,A}^{(N_f)}(x_i,x_j)$. For both kernel and $k$-point correlation function we have lightened the notation a bit, in suppressing or abbreviating the dependence on masses, temperature and topology. In the next step we have to determine the kernel. Before doing so, let us define the expectation value of an object $\mathcal{O}(\{x\})$ that only depends on the singular values of $W^\prime$
\begin{equation}
\mathbb{E}_{\mathcal{P}_{N,\nu}^{N_f}}\left[ \mathcal{O}(\{x\})\right]\equiv \frac{1}{Z_{N,\nu}^{(N_f)}(\{m\};\{a\})}\int_{\mathbb{R}_+^N}d[x]\ \mathcal{O}(\{x\})\ \mathcal{P}_{N,\nu}^{N_f}(\{x\},\{a\}).
\end{equation}
We once again only keep the dependence on the number of flavours,
topology and dimension $N$. Our main example for such expectation values will be over products and ratios of characteristic polynomials, defined in terms of singular values as 
\begin{equation}\label{DNdef}
D_N(z)\equiv\prod_{j=1}^N (z-x_j).
\end{equation}

We start with the computation of the kernel in the simpler case of the orthogonal polynomial ensemble \eqref{PNf0} at $A=0$. Here, the kernel is given by the orthogonal polynomials that we take in monic normalisation $p_k^{(N_f)}(x)=x^k+\mathcal{O}(x^{k-1})$. They are orthogonal with respect to some weight function
\begin{equation}
\label{OPdef}
\int_{\mathbb{R}_+}dx\ w^{(N_f)}(x) \ p_k^{(N_f)}(x) \ p_l^{(N_f)}(x)=\delta_{k,l}\ h_k^{(N_f)},
\end{equation}
which in our case is $w^{(N_f)}(x)=x^\nu e^{-x}\prod_{f=1}^{N_f}(x_j+m_f^2)$. The kernel then reads
\begin{equation}\label{kernel+w}
K_{N,0}^{(N_f)}(x,y)=\sqrt{w^{(N_f)}(x) \ w^{(N_f)}(y)}\ \widetilde{K}_{N,0}^{(N_f)}(x,y),
\end{equation}
where
\begin{equation}\label{KNsum}
\widetilde{K}_{N,0}^{(N_f)}(x,y)=\sum_{j=0}^{N-1}\frac{1}{h_j^{(N_f)}}p_j^{(N_f)}(x)p_j^{(N_f)}(y)
=\frac{1}{h_{N-1}^{(N_f)}}\frac{p_{N}^{(N_f)}(x)p_{N-1}^{(N_f)}(y)-p_{N}^{(N_f)}(y)p_{N-1}^{(N_f)}(x)}{x-y}.
\end{equation}
In the second step we have given the Christoffel-Darboux formula for the kernel valid at $x\neq y$. The reason why we have split off the weight functions  (which are automatically included in the kernel of the polynomial ensemble below), is that the  polynomial part of the kernel $\widetilde{K}_{N,0}^{(N_f)}(x,y)$ has a direct representation in terms of the expectation value  of two characteristic polynomials, see \eqref{KNdet2} below. 
In principle, the $p_k^{(N_f)}(x)$ can be constructed from a Gram-Schmidt procedure. They are in fact well known for $N_f=0$ and can be expressed via generalised Laguerre polynomials,
\begin{equation}\label{Laguerre}
p_k^{(N_f=0)}(x)=(-1)^kk!L_k^{\nu}(x),\quad h_k^{(0)}=k!\Gamma(k+\nu+1),
\end{equation}
in monic normalisation, including their norms. However, there is a much more elegant and direct construction of the kernel (and polynomials) for $N_f>0$, expressing them in terms of the quenched (Laguerre) polynomials with $N_f=0$. It was first observed in \cite{PZJ} that in any orthogonal polynomial ensemble the kernel can be expressed as the following expectation value of two characteristic polynomials:\footnote{Notice that also the orthogonal polynomials $p_k^{(N_f)}$ themselves can be expressed in terms of the expectation of a single $D_k$, the so-called Heine formula, cf. \cite{GA16}.}
\begin{eqnarray}\label{KNdet2}
\widetilde{K}_{N,0}^{(N_f)}(x,y)&=&\frac{1}{h_{N-1}^{(N_f)}}\mathbb{E}_{\mathcal{P}_{N-1,\nu}^{N_f}}\left[D_{N-1}(x)D_{N-1}(y)\right] 
\nonumber\\
\label{KNreweight}
&=&\frac{(-1)^{N_f}}{h_{N-1}^{(0)}}\frac{\mathbb{E}_{\mathcal{P}_{N-1,\nu}^{0}}\left[D_{N-1}(x)D_{N-1}(y)\prod_{f=1}^{N_f}D_{N-1}(-m_f^2)\right] }{\mathbb{E}_{\mathcal{P}_{N,\nu}^{0}}\left[\prod_{f=1}^{N_f}D_{N}(-m_f^2)\right]}.
\end{eqnarray}
Here, we have used the fact that
\begin{equation}\label{ZNhj}
Z_{N,\nu}^{(N_f)}(\{m\})=N!\prod_{j=0}^{N-1}h_j^{(N_f)},
\end{equation}
which follows from the theory of orthogonal polynomials, as well as 
\begin{eqnarray}
\frac{Z_{N,\nu}^{(N_f)}(\{m\})}{Z_{N,\nu}^{(0)}}&=&(-1)^{NN_f}\prod_{g=1}^{N_f}m_g^\nu\ \mathbb{E}_{\mathcal{P}_{N,\nu}^{0}}\left[\prod_{f=1}^{N_f}D_{N}(-m_f^2)\right]
\nonumber\\
&=&\prod_{f=1}^{N_f}m_f^\nu\ \Gamma(f+N)\frac{\det[L_{N+j-1}^\nu(-m_i^2)]_{i,j=1}^{N_f}}{\Delta_{N_f}(\{m^2\})},
\label{ZNvev}
\end{eqnarray}
see e.g. \cite{GA16} for a derivation. In other words, we have included the mass terms into the expectation value in \eqref{KNreweight} which is also called reweighting. Now everything is given in terms of quenched quantities at $N_f=0$, and we express the two expectation values in \eqref{KNreweight} in terms of determinants of generalised Laguerre polynomials, see \cite{GA16}
\begin{eqnarray}
\widetilde{K}_{N,0}^{(N_f)}(x,y)&=&
\frac{(-1)^{N_f-1}(N+N_f)!}{\Gamma(N+\nu)(y-x)\prod_{f=1}^{N_f}(y+m_f^2)(x+m_f^2)}
\frac{
\left|
\begin{array}{lll}
L_{N-1}^\nu(-m_1^2)&\ldots&L_{N+N_f}^\nu(-m_1^2)\\
\vdots& & \vdots\\
L_{N-1}^\nu(-m_{N_f}^2)&\ldots&L_{N+N_f}^\nu(-m_{N_f}^2)\\
L_{N-1}^\nu(x)&\ldots&L_{N+N_f}^\nu(x)\\
L_{N-1}^\nu(y)&\ldots&L_{N+N_f}^\nu(y)\\
\end{array}
\right|
}{\det[L_{N+g-1}^\nu(-m_f^2)]_{f,g=1}^{N_f}}.\nonumber\\
\label{Kernelmassive}
\end{eqnarray}
At $N_f=0$ this reduces to the standard Christoffel-Darboux form of the quenched kernel in terms of Laguerre polynomials \eqref{Laguerre}, which is well known.

For a similar construction of the unquenched polynomials $p_k^{(N_f)}(x)$, given  in terms of a ratio of determinants of Laguerre polynomials of size $N_f+1$ respectively $N_f$, we refer to \cite{DN, WGW}, where the Christoffel Theorem is used. Applying the Christoffel Theorem a second time to the kernel in Christoffel-Daboux form gives the same result as in \eqref{Kernelmassive}, as was observed in \cite{DN}.

Let us turn to the kernel in the polynomial ensemble \eqref{PNfA} with $A\neq0$.  It was shown for general polynomial ensembles in \cite{DF} that the kernel can be computed from the expectation value of the ratio of two characteristic polynomials as
\begin{equation}
\label{kernelgeneral}
K_{N,A}^{(N_f)}(x,y) = \frac{1}{x-y} \underset{z=y}{\Res}\ \mathbb{E}_{\mathcal{P}_{N,\nu}^{N_f}} \left[\frac{D_N(x)}{D_N(z)} \right] .
\end{equation}
Before we further exploit this relation, let us comment on the difference to \eqref{KNdet2}. 
For polynomial ensembles that are not of orthogonal polynomial type, relation \eqref{KNdet2} in terms of the product of two characteristic polynomials no longer holds, for the following reason: To derive \eqref{KNdet2}, each of the two characteristic polynomials can be included into one Vandermonde determinant $\Delta_N$, to form a larger one $\Delta_{N+1}$. Replacing each $\Delta_{N+1}$ by a determinant of orthogonal polynomials, expanding these and using orthogonality yields the sum in \eqref{KNsum}. However, if one of the determinants is no longer of Vandermonde type, this argument breaks down. Instead, roughly speaking, the residue fixes one of the arguments inside the non-Vandermonde determinant, that includes the Bessel-function in our case, and after further manipulation this results into the kernel as the sum of bi-orthogonal functions, one set of which are polynomials. We refer to \cite{DF} for a detailed derivation of \eqref{kernelgeneral}.

In \cite{ASW}, based on \eqref{kernelgeneral} a double integral formula was derived for the kernel for a certain class of polynomial ensembles, that includes our joint density \eqref{PNfA} for $N_f=0$, see \cite[Example 2.8]{ASW}. The same example was considered earlier in \cite{DF}. 
Let us spell out this quenched case first, before we turn to the inclusion of $N_f$ flavours, using again reweighting. 
We start with the first relation, 
\begin{equation}
\label{GmatrixentriesLaguerre}
g_{k,l}\equiv\int_0^{\infty} dx \ x^{k-1} \left(\frac{x}{a_l}\right)^{\nu/2} e^{-x-a_l} I_\nu (2\sqrt{a_l x}) =(k-1)! \ L_{k-1}^\nu(-a_l), 
\end{equation}
for $k=1,2,\ldots$, that spells out the matrix elements $g_{kl}$ of the Gram matrix $G=(g_{k,l})_{k,l=1}^{N}$ in term of the orthogonal polynomials we find at $A=0$. We note in passing that due to the Andr\'ei\'ef identity this immediately determines the partition function,
\begin{equation}\label{Zgram}
Z_{N,\nu}^{(0)}(\{a\})=N!\det[g_{k,l}]_{k,l=1}^N.
\end{equation}
Second, it holds that 
\begin{equation}
z^{k-1}= \int_{-\infty}^0ds (-1)^\nu \left(\frac{s}{z}\right)^{\nu/2}e^{s+z}I_{\nu}(2\sqrt{sz})(k-1)!L_{k-1}^\nu(-s)\ , \quad \mbox{for}\  k=1,2,\ldots.
\end{equation}
This means the integral over the Laguerre polynomials times the (analytically continued) generalised weight from the second determinant in the jpdf \eqref{PNfA} yields back monomial powers. Ensembles with this property are called  \emph{invertible polynomial ensembles}, and the resulting kernel follows from \cite[Proposition 2.10]{ASW}. 
For completeness we also give the expectation value of the ratio of two characteristic polynomials in this framework:
\begin{eqnarray}
\mathbb{E}_{\mathcal{P}_{N,\nu}^0} \left[\frac{D_{N}(x)}{D_{N}(z) } \right]
&=& \int_0^{\infty} dv \left(\frac{v}{z}\right)^{N-1}
\frac{(x-v)}{(z - v)}\int_0^{\infty} dt \left(\frac{t}{x}\right)^{\nu/2} e^{-t+x} J_\nu(2\sqrt{xt})
\prod_{n=1}^N(t+a_{n})
\nonumber\\
&&\times
\oint_{C} \frac{du}{2\pi i} \frac{-(v/u)^{\nu/2} e^{-u-v}I_\nu(2\sqrt{uv})}{(t+u)\prod_{n=1}^{N} (a_n-u)}.
\label{EratioNf0}
\end{eqnarray}
The integration contour $C$ encircles the points $a_1,\ldots,a_N$ counter-clockwise and leaves the real number $-t$ outside. Taking the residue as spelled out in \cite[Proposition 2.10]{ASW} leads to
\begin{equation} 
\label{kernelzero}
K_{N,A}^{(0)}(x,y) =\left(\frac{y}{x}\right)^{\nu/2} e^{x-y} \int_0^{\infty} dt t^{\nu/2} e^{-t}J_\nu(2\sqrt{tx}) \prod_{n=1}^{N} (t+a_n)
\oint_C  \frac{du }{2\pi i}   \frac{-u^{-\nu/2} e^{-u}I_\nu (2\sqrt{uy})}{ (t+u)\prod_{n=1}^{N} (a_n-u )} ,
\end{equation}
see \cite{DF} for an independent derivation.
Notice that when computing correlation functions from \eqref{Rk-kernel}, the pre-factor in front of the integral (also called cocycles) will drop out in all $k$-point correlation functions. Two kernels, which are related in this way, give the same correlation functions and are thus  called equivalent.

We are now ready to introduce $N_f$ flavours in our polynomial ensemble. Following the same idea of reweighting as in \eqref{KNreweight}, we can rewrite the ratio of  characteristic polynomials in \eqref{kernelgeneral} as 
\begin{equation}
\label{reweight2}
\mathbb{E}_{\mathcal{P}_{N,\nu}^{N_f}}\left[\frac{D_{N}(x)}{D_{N}(z)}\right] =
\frac{\mathbb{E}_{\mathcal{P}_{N,\nu}^{0}}\left[\frac{D_{N}(x)}{D_{N}(z)}\prod_{f=1}^{N_f}D_{N}(-m_f^2)\right] }{\mathbb{E}_{\mathcal{P}_{N,\nu}^{0}}\left[\prod_{f=1}^{N_f}D_{N}(-m_f^2)\right]}.
\end{equation}
Obviously, the expectation value in the denominator can be taken out of the residuum in \eqref{kernelgeneral}, allowing us to focus on denominator and numerator separately.
Both objects have been determined for our class of polynomial ensembles in \cite[Section 5]{ASW} as well, and so we can be brief. For the product of characteristic polynomials it holds
\begin{equation}
\mathbb{E}_{\mathcal{P}_{N,\nu}^{0}}\left[\prod_{f=1}^{N_f}D_{N}(-m_f^2)\right]
=(-1)^{N_f(N-\nu)}\prod_{f=1}^{N_f}m_f^{-\nu}e^{-m_f^2}\frac{\det[B_i(m_j)]_{i,j=1}^{N_f}}{\Delta_{N_f}(\{m^2\})},
\label{EprodD}
\end{equation}
where we have defined 
\begin{equation}
\label{Bdef}
B_i(m^2)\equiv\int_0^\infty dt \ t^{i-1+\nu/2}e^{-t}I_\nu(2m\sqrt{t})\prod_{n=1}^N(t+a_n).
\end{equation}
The ratio we need in \eqref{KNreweight} is given by 
\begin{eqnarray}
&&
\mathbb{E}_{\mathcal{P}_{N,\nu}^0} \left[ \frac{D_N(x_1)}{D_N(z)} \prod_{f=1}^{N_f} D_N(-m_f^2) \right]
\nonumber\\
&=&
\frac{(-1)^{N_f(N+\nu)}x^{-\nu/2}e^x\prod_{f=1}^{N_f}m_f^{-\nu}e^{-m_f^2}}{\prod_{g=1}^{N_f}(x+m_g^2)\Delta_{N_f}(\{m^2\})}
\int_0^\infty dv \left(\frac{v}{z}\right)^{N-1}\frac{x-v}{z-v}\prod_{f=1}^{N_f}(m_f^2+v)
\nonumber\\
&&\times\oint_{C} \frac{du}{2\pi i} \frac{(v/u)^{\nu/2}e^{-u-v} }{\prod_{n=1}^{N} (a_n-u)}
I_\nu(2\sqrt{uv})
\det \left[ \begin{matrix}
 \hat{A}(x,u)&A(m_{1}^2,u) & \ldots & A(m_{N_f}^2,u) \\
\hat{B}_{1}(x)&B_{1}(m_1^2) & \ldots &B_{1}(m_{N_f}^2)  \\
\vdots &\vdots & \ldots & \vdots \\
\hat{B}_{N_f}(x)&B_{N_f}(m_1^2) & \ldots & B_{N_f}(m_{N_f}^2) \\
\end{matrix} \right].
\label{ratioLaguerre}
\end{eqnarray}
Here, we have defined further quantities,
\begin{equation}
\label{Bhatdef}
\hat{B}_i(x)\equiv\int_0^\infty dt \ t^{i-1+\nu/2}e^{-t}J_\nu(2\sqrt{xt})\prod_{n=1}^N(t+a_n),
\end{equation}
where the hat $\ \widehat{} \ $ indicates that the integral contains a Bessel-$J$ function,
\begin{equation}\label{Bessel-Jdef}
J_\nu(z)=\sum_{n=0}^\infty \frac{(-1)^n(z/2)^{2n+\nu}}{n!\Gamma(n+\nu+1)},
\end{equation}
instead of the modified Bessel-$I$ compared to \eqref{Bdef}. Likewise we define
\begin{eqnarray}
\label{Ahatdef}
\hat{A}(x,u) &\equiv& \int_0^\infty dt \ t^{\nu/2}e^{-t}J_\nu(2\sqrt{xt})\frac{-1}{u+t}\prod_{n=1}^N(t+a_n),\\
A(m^2,u) &\equiv& \int_0^\infty dt \ t^{\nu/2}e^{-t}I_\nu(2m\sqrt{t})\frac{-1}{u+t}\prod_{n=1}^N(t+a_n).
\label{Adef}
\end{eqnarray}
If we insert the results \eqref{EprodD} and \eqref{ratioLaguerre} into \eqref{reweight2}, we can now take the residuum in \eqref{kernelgeneral} for an arbitrary number of flavours. The residuum picks the pole at $z=v$ inside the integral in the first line on the right hand side of \eqref{ratioLaguerre}, and we are led to the following result:
\begin{eqnarray}
K_{N,A}^{(N_f)}(x,y) &=& \frac{x^{-\nu/2}e^{x}}{\det[B_i(m_j)]_{i,j=1}^{N_f}}\prod_{f=1}^{N_f}\frac{y+m_f^2}{x+m_f^2}
\oint_{C} \frac{du}{2\pi i} \frac{(y/u)^{-\nu/2}e^{-u-y} }{\prod_{n=1}^{N} (a_n-u)}
I_\nu(2\sqrt{uy})
\nonumber\\
&&\times
\det \left[ \begin{matrix}
 \hat{A}(x,u)&A(m_{1}^2,u) & \ldots & A(m_{N_f}^2,u) \\
\hat{B}_{1}(x)&B_{1}(m_1^2) & \ldots &B_{1}(m_{N_f}^2)  \\
\vdots &\vdots & \ldots & \vdots \\
\hat{B}_{N_f}(x)&B_{N_f}(m_1^2) & \ldots & B_{N_f}(m_{N_f}^2) \\
\end{matrix} \right]
\nonumber\\
&=& 
\prod_{f=1}^{N_f}\frac{y + m_f^2}{x + m_f^2} \ 
\frac{
\det \left[ \begin{matrix}
K_{N,A}^{(0)}(x,y) & \hat{K}_{N,A}^{(0)}(m_1^2,y) & \ldots & \hat{K}_{N,A}^{(0)}(m_{N_f}^2,y) \\
\hat{B}_1(x)& B_1(m_1^2) & \ldots & B_1(m_{N_f}^2) \\
\vdots & \ldots & \ldots & \vdots \\
 \hat{B}_{N_f}(x)&B_{N_f}(m_1^2) & \ldots & B_{N_f}(m_{N_f}^2) \\
\end{matrix} \right] }{\det[B_{i}(m_f^2)]_{i,f=1}^{N_f} }.
\nonumber\\
\label{kernel-final}
\end{eqnarray}
In the second step we have pulled the $u$-integral into the first row of the determinant, giving rise to the quenched kernel \eqref{kernelzero}, as well as to its variant  with a Bessel-$J$ replaced by a Bessel-$I$ function,
\begin{equation}\label{hatKNA}
\hat{K}_{N,A}^{(0)}(m^2,y) \equiv\left(\frac{y}{x}\right)^{\nu/2} e^{x-y} \int_0^{\infty} dt t^{\nu/2} e^{-t}I_\nu(2m\sqrt{t}) \prod_{n=1}^{N} (t+a_n)
\oint_C  \frac{du }{2\pi i}   \frac{-u^{-\nu/2} e^{-u}I_\nu (2\sqrt{uy})}{ (t+u)\prod_{n=1}^{N} (a_n-u )} .
\end{equation}
Once again, the ratio in front of the determinant \eqref{kernel-final} as well as the prefactors in front of the integrals in the kernels \eqref{kernelzero} and \eqref{hatKNA} can be dropped (after taking the latter out of the determinant), constituting a kernel equivalent to $K_{N,A}^{(N_f)}(x,y)$ in \eqref{kernel-final}. This is the starting point for the asymptotic large-$N$ analysis in the next section.

\sect{The microscopic large-$N$ limit at the origin}
\label{sec:asymptotic}

In this section we will investigate the microscopic Dirac operator spectrum at non-zero temperature for an arbitrary number of quark flavours at fixed topology $\nu$. This will be done by a saddle point analysis of the double integral representation of the quenched kernel \eqref{kernelzero} as a first step in Subsection \ref{subsec:qK}, and of the remaining building blocks of the unquenched kernel \eqref{kernel-final} in Subsection \ref{subsec:uqK}. This yields all $k$-point correlation functions in this limit. A comparison to the known unquenched zero temperature results and resulting universality statement will be the subject of Section \ref{sec:Universal}.

Starting from the model \eqref{ZNA}, we have so far worked with $N$-independent weight functions in \eqref{PNfA}. This was done for technical reasons, e.g. to be able to relate expectation values of characteristic polynomials of different sizes. In order to obtain a compact support for the limiting macroscopic density
of the squared singular values of $W^\prime$, we rescale the $x_j\to N x_j$, and accordingly the $a_j\to N a_j$. This leads to a factor of $N$ in the exponent of \eqref{Zshift}. Often an additional scale is introduced in the QCD literature, namely the chiral condensate $\Sigma$ at zero temperature. 
In the Banks-Casher relation it appears as $V\Sigma$ together with the volume $V$, which is replaced here by $N\Sigma$ instead. 
We will set the parameter $\Sigma$ to unity and will measure the eigenvalues on the scale of the temperature dependent condensate $\Xi$ to be introduced later. 
The smallest Dirac operator eigenvalues then live on the microscopic scale $1/N$ in the vicinity of the origin, also called the hard edge in random matrix theory. 

\subsection{The quenched limiting kernel at non-zero temperature}
\label{subsec:qK}
We begin with the quenched case $N_f=0$, to consider the following rescaled kernel
\begin{equation}
\label{k-def}
k(\rho, \eta) = \frac{1}{N} K_{N,A}^{(0)}\left(x_1=\frac{\rho}{N},x_2= \frac{\eta}{N}\right).
\end{equation}
In order to get an expression from \eqref{kernelzero} that has a limit (and to keep a meaningful integration contour $C$) we also rescale the integration variables $t\to Nt$ and $u\to Nu$. The rescaled kernel then takes the form
\begin{equation}
\label{kernelrescaled}
k(\rho,\eta) =\int_0^{\infty} dt\ t^{\nu/2} J_\nu\left(2\sqrt{\rho t}\right) e^{-N\mathcal{L}_2(t)} \oint_C \frac{du}{2\pi i} \ \frac{-u^{-\nu/2} I_\nu\left(2\sqrt{\eta u}\right)}{s+u} e^{-N\mathcal{L}_1(u)},
\end{equation}
where we have defined
\begin{eqnarray}
\mathcal{L}_1(u) &=& u+ \frac{1}{N} \sum_{n=1}^{N} \log(a_n-u),
\nonumber\\
\mathcal{L}_2(t) &=& t- \frac{1}{N} \sum_{n=1}^{N} \log(a_n+t).
\label{L12def}
\end{eqnarray}
These result from rewriting
\begin{eqnarray}
\frac{e^{-Nu}}{\prod_{n=1}^{N} (a_n-u)}&=& e^{-N\mathcal{L}_1(u)},
\nonumber\\
e^{-Nt} \prod_{n=1}^{N} (t+a_n) &=& e^{-N\mathcal{L}_2(t)},
\end{eqnarray}
after rescaling all variables. 
Clearly the leading contribution to the rescaled kernel \eqref{kernelrescaled} will come from the saddle points of $\mathcal{L}_1(u)$ and $\mathcal{L}_2(t)$.
Apart from $t\in\mathbb{R}_+$ and $u\in C$ the two are closely related. We have for $\mathcal{L}_1(u)$
\begin{eqnarray}
\mathcal{L}_1^\prime (u) &=& 1 - \frac{1}{N} \sum_{n=1}^{N} \frac{1}{a_n-u} ,
\nonumber\\
\mathcal{L}_1^{\prime \prime} (u) &=& - \frac{1}{N} \sum_{n=1}^{N} \frac{1}{(a_n-u)^2} ,
\label{SPL1}
\end{eqnarray}
and for a real variable $x\neq\pm a_l$ for all $l=1,\ldots,N$ the following simple relations hold:
\begin{eqnarray}
\mathcal{L}_1(-x) &=& -\mathcal{L}_2(x) ,
\nonumber\\
\mathcal{L}_1^{\prime}(-x) &=& +\mathcal{L}_2^{\prime}(x) ,
\nonumber\\
\mathcal{L}_1^{\prime \prime}(-x) &=& -\mathcal{L}_2^{\prime \prime}(x).
\label{L12rel}
\end{eqnarray}
Thus if $\bar{t}$ is a saddle point of $\mathcal{L}_2$ then $\bar{u}=-\bar{t}$ is a saddle point $\mathcal{L}_1$. Obviously we will have to deform the integration contour to reach the saddle point. Before describing the outcome let us remark the following: 
In \cite{GuhrWettig} a similar representation of the quenched kernel \eqref{kernelrescaled} was derived, either as a double real integral or as an integral over a graded $2\times2$ matrix, resulting from the supersymmetric method applied there. The saddle point analysis is remarkably similar, see \cite[Appendix C]{GuhrWettig}, and we will follow it quite closely. However, to be self-contained we will give some details here as well.

An important role is played by the critical value $t_c\in\mathbb{R}_+$ defined as 
\begin{equation}
t_c \equiv  \frac{1}{N} \sum_{n=1}^{N} \frac{1}{a_n}>0,
\label{tcdef}
\end{equation}
where we recall that $a_n>0$ for all $n$.
Now we analyse the saddle points of $\mathcal{L}_2(t)$. Consider the derivative $\mathcal{L}^\prime_2(t)$ given by the function
\begin{equation}
h(t) = 1-\frac{1}{N}\sum_{n=1}^{N} \frac{1}{a_n+t},
\label{Hdef}
\end{equation}
which is defined on $\mathbb{R}\backslash\Omega$, where $\Omega = \{ -a_1,\ldots, -a_N \}$. 
Its derivative $\mathcal{L}^{\prime \prime}_2(t)$ is always positive:
\begin{equation}
h^\prime(t) = \frac{1}{N} \sum_{n=1}^{N} \frac{1}{(a_n+t)^2} > 0, \forall \ t \in \mathbb{R}\backslash \Omega.
\label{hprime}
\end{equation}
For $t\rightarrow \pm \infty$ we have $h(t) \rightarrow 1$. 
Furthermore, $h(t)$ has $N$ real, negative poles at the points $-a_1,\ldots, -a_N$. For $t\in [0,\infty]$ the function $h(t)$ is thus smooth, continuous and monotonically increasing with limit $1$ as $t\rightarrow \infty$. Also, $h(t)$ has $N$ zeros in total on $\mathbb{R}$. Because $h(t)$ has $N$ real, negative poles, $N-1$ zeros must be real and negative, located between the poles. The location of the remaining zero is then determined by the value of $t_c$:
\begin{equation}
h(0) = 1-t_c.
\end{equation}
For $t_c>1$ we have $h(0)<0$ and thus a unique zero $\bar{t}>0$ exists, $h(\bar{t})=0$, which is in the domain of integration and constitutes our saddle point, due to \eqref{hprime}. In contrast, for $0<t_c<1$ the zero is located outside the domain of integration, and it can be shown in analogy to 
\cite[Appendix B]{GuhrWettig} that when integrating over the fluctuations the integral vanishes. The critical case $t_c=1$ corresponds to the case when the temperature dependent chiral condensate is vanishing \cite{WSW}. In the following we will only consider the subcritical case $t_c>1$.
In analogy to \cite{GuhrWettig} we will denote the solution of the saddle point equation for $\mathcal{L}_2(t)$ by $\bar{t}$:
\begin{equation}
0 = 1-\frac{1}{N}\sum_{n=1}^{N} \frac{1}{a_n+\bar{t}}\ ,
\label{SP}
\end{equation}
where the temperature dependent chiral condensate is then
\begin{equation}
\Xi=\Xi(\{a\})=\bar{t}.
\end{equation}
$\Xi$ will set the scale, and in a proper, refined scaling of \eqref{k-def} the eigenvalues of the Dirac operator will be rescaled by it, see \eqref{finalkernel0} below. This leads to a parameter free prediction from the random matrix ensemble.

The analysis of $\mathcal{L}_1(u)$ is similar. We introduce the function
\begin{equation}
g(u) = 1- \frac{1}{N} \sum_{n=1}^{N} \frac{1}{a_n - u},
\end{equation}
which is the derivative $\mathcal{L}^\prime_1(u)$. Since the contour $C$ encircles the points $\widetilde{\Omega}= \{a_1,\ldots, a_N \}$ and leaves $-t$ outside, $g(u)$ is an analytic function on the contour $C$ and we can deform $C$ as long as we do not cross the singularities which are the points in $\widetilde{\Omega}$. In addition, we have
\begin{equation}
g(0) = 1-t_c <0,
\end{equation}
due to $t_c>1$ being subcritical, and
\begin{equation}
g^\prime(u) = -\frac{1}{N} \sum_{n=1}^{N} \frac{1}{(a_n-u)^2} <0, \quad \forall \ u \in \mathbb{R}\backslash \widetilde{\Omega}.
\end{equation}
For real $u$ our function $g(u)$ is continuous and decreases monotonically with $\lim_{u\rightarrow \pm \infty} g(u) =1$. Since $g(u)$ is analytic and continuous in $u$ with $N$ real poles, without loss of generality the contour $C$ can be deformed such that the zeros of $g(u)$ are also real.\footnote{One can also use the fundamental theorem of algebra to show that $g(u)$ has only real zeros.} Because of the continuity of $g(u)$ and the positioning of the poles, there are exactly $N-1$ real, positive zeros and one real, negative zero. This zero corresponds to a maximum of $\mathcal{L}_1(u)$ instead of the required minimum, and we will return to this point below.

In view of the relation between the saddle points on $\mathcal{L}_1(u)$ and $\mathcal{L}_2(t)$, $\bar{u} = - \bar{t}$, we have not only to expand around the saddle points to compute the fluctuations, but also to take care of the resulting pole  
$1/(u+t)$ in \eqref{kernelrescaled} that couples the two integrals. For the real positive variable $t$ we thus write
\begin{equation}
t = \bar{t} + \frac{1}{\sqrt{N}} \ x. 
\end{equation}
Furthermore, the saddle point at $\bar{u}$ is not a minimum. For complex $u$ with a suitable branch cut of the logarithm in $\mathcal{L}_1(u)$ we can expand via
\begin{equation}
u = \bar{u} + i \frac{1}{\sqrt{N}} \ y ,
\end{equation}
which means that the resulting integration over $y$ gets rotated to the real line as $N \rightarrow \infty$ and the maximum we obtained at $\mathcal{L}_1(\bar{u})$ becomes a minimum as required for the saddle point expansion.

We begin with the evaluation of the contour integral, denoted by 
\begin{equation}
\label{Iintdef}
\mathcal{I}(t) = \oint_C \frac{du}{2\pi i} \ \frac{f_1(u)}{u+t} e^{-N\mathcal{L}_1(u)},
\end{equation}
with $f_1(u)=u^{-\nu/2} I_\nu(2\sqrt{\eta u})$.
First, following the strategy of \cite{FGS},
we deform the contour $C\to\widetilde{C}$ to pass through the saddle point $\bar{u}$. This adds a contribution from the pole at $-t$ if and only if $\bar{u}<-t$:
\begin{eqnarray}
\mathcal{I}(t)&=&  \oint_{\widetilde{C}} \frac{du}{2\pi i} \ \frac{f_1(u)}{u+t} e^{-N\mathcal{L}_1(u)} - \oint_{C_t} \frac{du}{2\pi i}  \ \Theta(-t-\bar{u}) \frac{f_1(u)}{u+t} e^{-N\mathcal{L}_1(u)}
\nonumber\\
&\approx&  
\frac{1}{2\pi i} \int_{-\infty}^{\infty} \frac{dy}{\sqrt{N}} \ \frac{f_1(-\bar{t})}{\bar{u}+iN^{-1/2}y+t} e^{N\mathcal{L}_2(\bar{t})-\frac{1}{2}\mathcal{L}_2^{\prime \prime}(\bar{t})y^2} 
 -\Theta(-t+\bar{t})f_1(-t)e^{N\mathcal{L}_2(t)},\quad\quad
\end{eqnarray}
where $\Theta(x)$ denotes the Heaviside or step function. 
In the first integral we have expanded around the saddle point,
\begin{equation}
-N\mathcal{L}_1(u) = -N \mathcal{L}_1(\bar{u}) + \frac{1}{2} \mathcal{L}_1^{\prime \prime}(\bar{u}) y^2 + \mathcal{O}(N^{-1/2}) ,
\end{equation}
and used \eqref{L12rel}. 
Likewise, in the second integral we have evaluated the contour integral over $C_t$ around the pole at $u=-t$. 
To this order the rescaled kernel \eqref{kernelrescaled} thus takes the form
\begin{eqnarray}
k(\rho,\eta) &\approx& \int_0^{\infty} dt \ f_2(t) e^{-N\mathcal{L}_2(t)} f_1(-t) e^{N\mathcal{L}_2(t)} \Theta(\bar{t}-t) 
\nonumber\\
&&-\frac{1}{2\pi \sqrt{N}} \int_{-\infty}^{\infty} dy \ e^{-\frac{1}{2} \mathcal{L}_2^{\prime \prime}(\bar{t}) y^2} \int_0^{\infty} dt \ f_2(t) e^{-N\mathcal{L}_2(t)} e^{N\mathcal{L}_2(\bar{t})} \frac{f_1(-\bar{t}) }{-\bar{t}+iN^{-1/2}y+t},
\end{eqnarray}
where we have introduced $f_2(t) = t^{\nu/2} J_\nu(2\sqrt{\rho t})$.
In the first line the exponents cancel and the integral truncates. In the second line we also have to expand around the saddle point,
\begin{equation}
-N\mathcal{L}_2(t)=-N \mathcal{L}_2(\bar{t}) - \frac{1}{2} \mathcal{L}_2^{\prime \prime}(\bar{t}) x^2 + \mathcal{O}(N^{-1/2}).
\end{equation}
This leads to the following expression
\begin{equation}
k(\rho,\eta) \approx\int_0^{\bar{t}} dt \ f_2(t) f_1(-t) -\frac{f_2(\bar{t})f_1(-\bar{t})}{2\pi \sqrt{N}} \int_{-\infty}^{\infty} dy \ e^{-\frac{1}{2} \mathcal{L}_2^{\prime \prime}(\bar{t}) y^2}\int_{-\infty}^{\infty} \frac{dx}{\sqrt{N}} \   e^{-\frac{1}{2} \mathcal{L}_2^{\prime \prime}(\bar{t}) x^2} \frac{\sqrt{N}}{x+iy} .
\end{equation}
Clearly, the integrals in the second term converge and are finite, and thus this term is suppressed by $\mathcal{O}(N^{-1/2})$. 
The final answer in the first integral can be evaluated:
\begin{eqnarray}
\int_0^{\bar{t}} dt \ f_1(-t) f_2(t) &=&\bar{t} \int_0^{1} d\tau \  J_\nu(\sqrt{4\eta \bar{t}\tau}) J_\nu(\sqrt{4\rho \bar{t}\tau})
\nonumber\\
&=&2\bar{t} 
\frac{\sqrt{4\eta\bar{t}}J_{\nu+1}(\sqrt{4\eta\bar{t}})J_\nu(\sqrt{4\rho\bar{t}})-\sqrt{4\rho\bar{t}}J_{\nu+1}(\sqrt{4\rho\bar{t}})J_\nu(\sqrt{4\eta\bar{t}})}{4\bar{t}(\eta-\rho)},
\label{Bessel-integralJJ}
\end{eqnarray}
where the last line only holds for unequal arguments. A similar expression holds for equal arguments as shown below. In a final step we move from the Wishart eigenvalues $x_j$ of the matrix $WW^\dag$ to the Dirac eigenvaues $\zeta_j$, by simply squaring $x_j=\zeta_j^2$. This leads to a Jacobian $\sqrt{|2\zeta2\eta|}$ we have to multiply to the kernel as follows:
\begin{equation}
\mathcal{K}_A^{(0)}(\zeta,\eta)=\lim_{N\to\infty}
\frac{2\sqrt{|\zeta\eta|}}{4N \Xi} K_{N,A}^{(0)}\left(x=\frac{\zeta^2}{4N\Xi}, y= \frac{\eta^2}{4N \Xi} \right) =\sqrt{|\zeta\eta|}\ \frac{\zeta J_{\nu+1}(\zeta)J_\nu(\eta)-\eta J_{\nu+1}(\eta)J_\nu(\zeta)}{\zeta^2-\eta^2},
\label{finalkernel0}
\end{equation}
where we have dropped the cocycles in \eqref{kernelzero} before taking the limit.
This is the final result of this subsection. It equals the well-known universal Bessel-kernel for the quenched Dirac operator spectrum at zero temperature \cite{ADMN}, see also Subsection \ref{subsec:equiv}. This universality result for non-zero temperature in the subcritical regime, when measuring the Dirac operator eigenvalues (hence the squared arguments) in units of the temperature dependent condensate $\Xi=\bar{t}$, was derived previously in \cite{GuhrWettig} for $\nu=0$ using supersymmetry. In the next subsection we will derive a determinantal expression for the unquenched kernel at non-zero temperature which is universal as well, that was not previously known.

For completeness we also give the limiting result for the quenched microscopic density at non-zero temperature, the kernel at equal arguments, where we have 
\begin{equation}
\label{rhonu}
\mathcal{R}_{1,A}^{(0)}(\zeta)=\mathcal{K}_A^{(0)}(\zeta,\zeta)=\lim_{N\to\infty}
\frac{2|\zeta|}{4N \Xi} K_{N,A}^{(0)}\left(\frac{\zeta^2}{4N\Xi}, \frac{\zeta^2}{4N \Xi} \right) 
=\frac{|\zeta|}{2}\left( J_\nu^2(\zeta) - J_{\nu-1}(\zeta)J_{\nu+1}(\zeta)\right).
\end{equation}

\subsection{The unquenched limiting kernel at non-zero temperature}
\label{subsec:uqK}

We turn to the limit of the unquenched kernel given by \eqref{kernel-final}, where we have to determine the asymptotic of its building blocks. Fortunately, we can use the saddle point analysis of the previous subsection. 
Let us emphasise that in contrast to full QCD the number of flavours does not influence the saddle point solution, and that for any fixed $N_f$ we can be in the broken phase. 

Recalling the results from the previous subsection, in the second building block $\hat{K}_{N,A}^{(0)}(m^2,y)$ of the unquenched kernel given in \eqref{hatKNA}, we merely have to replace the Bessel-$J$ function with a Bessel-$I$, with argument $\rho\to\mu^2$,  compared to the quenched kernel. The saddle point evaluation thus equally goes through, and we obtain the following integral instead of \eqref{Bessel-integralJJ}:
\begin{equation}
\bar{t} \int_0^{1} d\tau \  I_\nu(\sqrt{4\mu^2\bar{t}\tau}) J_\nu(\sqrt{4\eta \bar{t}\tau})=
2\bar{t} 
\frac{\sqrt{4\mu^2\bar{t}}I_{\nu+1}(\sqrt{4\mu^2\bar{t}})J_\nu(\sqrt{4\eta\bar{t}})+\sqrt{4\eta\bar{t}}J_{\nu+1}(\sqrt{4\eta\bar{t}})I_\nu(\sqrt{4\mu^2\bar{t}})}{4\bar{t}(\mu^2+\eta)}.
\label{Bessel-integralIJ}
\end{equation}
This immediately leads us to the results
\begin{equation}
\mathcal{B}_{\rm IJ}(\mu,\zeta)=\lim_{N\to\infty}
\frac{2\sqrt{|\mu\zeta|}}{4N \Xi} \hat{K}_{N,A}^{(0)}\left(m^2=\frac{\mu^2}{4N\Xi}, y= \frac{\zeta^2}{4N \Xi} \right) =\frac{\mu I_{\nu+1}(\mu)J_\nu(\zeta)+\zeta J_{\nu+1}(\zeta)I_\nu(\mu)}{\mu^2+\zeta^2}.
\label{BIJdef}
\end{equation}
For convenience we also define this suggestive notation for the quenched kernel \eqref{finalkernel0} above:
\begin{equation}
\label{KBdef}
\mathcal{B}_{\rm JJ}(\zeta_a,\zeta_b)=\mathcal{K}_A^{(0)}(\zeta_a,\zeta_b).
\end{equation}

Next we have to determine the asymptotic of the determinant $\det[B_i(m_j^2)]_{i,j=1}^{N_f}$ in the denominator of the kernel $K_{N,A}^{(N_f)}(x,y)$ in \eqref{kernel-final}. We will use later that many of the constants, that appear in manipulating the 
matrix 
$B_i(m_j^2)$, can be taken out of the determinant and will cancel when doing the same manipulations on the determinant in the numerator of the kernel.

First, a rescaling of $t\to Nt$, $a_n\to Na_n$ and $m^2=\frac{\mu^2}{4N\bar{t}}$ upon \eqref{Bdef} leads to 
\begin{eqnarray}
B_i\left(\frac{\mu^2}{4N\bar{t}}\right)&=&N^{N+\frac{\nu}{2}+i}\int_0^\infty dtt^{i-1+\nu/2}I_\nu(\mu\sqrt{t/\bar{t}})e^{-N\mathcal{L}_2(t)}
\nonumber\\
&\approx& N^{N+\frac{\nu}{2}+i}e^{-N\mathcal{L}_2(\bar{t})}\int_{-\infty}^\infty
dx(\bar{t}+x/\sqrt{N})^{i-1+\nu/2}I_\nu\left(\mu\sqrt{1+x/(\bar{t}\sqrt{N})}\right)
e^{-\frac12\mathcal{L}_2^{\prime\prime}(\bar{t})x^2}\!,\quad\quad
\label{B1asym}
\end{eqnarray}
where we have expanded around the saddle point at $t=\bar{t}+x/\sqrt{N}$. We have kept the fluctuations in the prefactor of the exponential under the integral. This is necessary because of the following argument: Had we considered only  the leading order contribution at $x=0$, and taken these factors out of the integral, the determinant over the $B_i(m^2_j)$ would become highly degenerate and thus vanish. 

Next, we change variables, $x=p\sqrt{2/\mathcal{L}_2^{\prime\prime}(\bar{t})}$, split off the power $t^{i-1}$ and use that 
under the determinant we can bring $(\bar{t}+x/\sqrt{N})^{i-1}$ back to monic power in $\sim p^{i-1}$, when subtracting the appropriate multiples of upper rows. 
 We denote the resulting matrix by $\widetilde{B}_i(\frac{\mu_j^2}{4N\bar{t}})$. 
Dropping for now the constants in \eqref{B1asym}, we thus arrive at the following term in the determinant:
\begin{eqnarray}
\widetilde{B}_i\left(\frac{\mu^2}{4N\bar{t}}\right)&\sim&\sqrt{N}\sigma^i \int_{-\infty}^\infty dp\ (\bar{t}+\sigma p)^{\nu/2}
I_\nu\left(\mu\sqrt{1+\sigma p/\bar{t}}\right) e^{-p^2} p^{i-1}
\nonumber\\
&\sim&\sqrt{N}\sigma^i \int_{-\infty}^\infty dp\ (\bar{t}+\sigma p)^{\nu/2}
I_\nu\left(\mu\sqrt{1+\sigma p/\bar{t}}\right) \left(\frac{-1}{2}\right)^{i-1}\frac{d^{i-1}}{dp^{i-1}}e^{-p^2},
\end{eqnarray}
with the abbreviation $\sigma =\sqrt{2/(N\mathcal{L}_2^{\prime\prime}(\bar{t}))}$. In the second step, we have used the same argument 
in simplifying the matrix $\widetilde{B}_i(\frac{\mu_j^2}{4N\bar{t}})$ under the determinant:
to replace the monic powers $p^{i-1}$ by Hermite polynomials in monic normalisation,
$h_{i-1(}p)=H_{i-1}(p)/2^i=(-1/2)^{i-1}e^{p^2}\frac{d^{i-1}}{dp^{i-1}}e^{-p^2}$, and then applied the Rodrigues formula, as inspired by the derivation in \cite{FGS}. We can do a repeated integration by parts, where all boundary terms vanish. 
The differentiation is acting now on the Bessel-$I$ function, and its prefactor can be simplified by introducing the variable
$z=\mu\sqrt{1+\sigma p/\bar{t}}$. The resulting matrix is denoted by 
$\widetilde{\widetilde{B}}_i(\frac{\mu_j^2}{4N\bar{t}})$, 
 and we get 
\begin{eqnarray}
\widetilde{\widetilde{B}}_i\left(\frac{\mu^2}{4N\bar{t}}\right)&=& \frac{\sqrt{N}\sigma^i}{2^{i-1}}
\int_{-\infty}^\infty dp\left(\frac{\mu^2\sigma}{2\bar{t}}\right)^{i-1}
\left(\frac{d}{zdz}\right)^{i-1}\left[\frac{\bar{t}^{\nu/2}}{\mu^\nu}z^\nu I_\nu(z)\right]\Bigg|_{p=0}
e^{-p^2}\nonumber\\
&= & \frac{\sqrt{N}\sigma^i}{2^{i-1}}\left(\frac{\mu^2\sigma}{2\bar{t}}\right)^{i-1}\frac{\bar{t}^{\nu/2}}{\mu^\nu}\int_{-\infty}^\infty dp\ \left[z^{\nu-(i-1)} I_{\nu-(i-1)}(z)\right]\Big|_{p=0}\  e^{-p^2}\nonumber\\
&=& \sqrt{N} \left(\frac{\sigma}{2}\right)^{i-1} \bar{t}^{\frac{\nu}{2}-i+1}\mu^{i-1}I_{\nu-i+1}(\mu) \sqrt{\pi},
\end{eqnarray}
where in the pre-exponential factor we can set $p=0$, once we no longer obtain a degenerate determinant. The differentiation of the combination $z^\nu I_\nu(z)$ is well known, see e.g. \cite[10.29.4]{NIST}. In the last step we can set $p=0$ as now the degeneracy is completely lifted, which amounts to setting $z=\mu$, and do the remaining Gaussian integral. Overall, including all prefactors we have derived the following result:
\begin{eqnarray}
\det\left[B_i\left(\frac{\mu_j^2}{4N\bar{t}}\right)\right]_{i,j=1}^{N_f}&\approx&
\prod_{i=1}^{N_f}\left(N^{N+\frac{\nu+1}{2}+i}e^{-N\mathcal{L}_2(\bar{t})}\left(\frac{\sigma}{2}\right)^{i-1} \bar{t}^{\frac{\nu}{2}-i+1}\sqrt{\pi}\right)
\det\left[\mu_j^{i-1}I_{\nu-i+1}(\mu_j)\right]_{i,j=1}^{N_f}\nonumber\\
&\sim&\det\left[\mu_j^{i-1}I_{\nu+i-1}(\mu_j)\right]_{i,j=1}^{N_f}.
\label{detBs}
\end{eqnarray}
In the last step we have used the following identity \cite{NIST}
\begin{equation}
\mu I_{\nu+1}(\mu)=\mu I_{\nu-1}(\mu)-2\nu I_\nu(\mu),
\label{Bessel-Iid}
\end{equation}
so that by successive addition of multiples of upper rows to lower rows in the determinant the index in the Bessel-$I$ function is now increasing, rather than decreasing.

The determinant in the numerator of \eqref{kernel-final} has one extra row of kernels we have already analysed and one extra row with $\hat{B}_i(x)$ containing Bessel-$J$ functions, see \eqref{Bhatdef}. Rescaling $x=\frac{\zeta^2}{4N\bar{t}}$ in addition to $t$ and the $a_n$ we depart from 
\begin{eqnarray}
\hat{B}_i\left(\frac{\zeta^2}{4N\bar{t}}\right)&=&N^{N+\frac{\nu}{2}+i}\int_0^\infty dtt^{i-1+\nu/2}J_\nu(\zeta\sqrt{t/\bar{t}})e^{-N\mathcal{L}_2(t)}
\nonumber\\
&\approx& N^{N+\frac{\nu}{2}+i}e^{-N\mathcal{L}_2(\bar{t})}\int_{-\infty}^\infty
dx(\bar{t}+x/\sqrt{N})^{i-1+\nu/2}J_\nu\left(\zeta\sqrt{1+x/(\bar{t}\sqrt{N})}\right)
e^{-\frac12\mathcal{L}_2^{\prime\prime}(\bar{t})x^2}\!.\quad\quad
\end{eqnarray} 
Following the same steps as above, the same operations under the determinant as on the functions $B_i$ lead to the replacement of $t^{i-1}$ by $p^{i-1}$ by $h_{i-1}(p)$. Defining $z=\zeta\sqrt{1+\sigma p/\bar{t}}$ instead, we can use the identity for Bessel-$J$ functions \cite[10.6.6]{NIST}
\begin{equation}
\left(\frac{d}{zdz}\right)^{i-1}\left[z^\nu J_\nu(z)\right]=z^{\nu-(i-1)} J_{\nu-(i-1)}(z),
\end{equation}
to arrive at the following replacement under the determinant:
\begin{equation}
\hat{B}_i\left(\frac{\zeta^2}{4N\bar{t}}\right)\to N^{N+\frac{\nu+1}{2}+i}e^{-N\mathcal{L}_2(\bar{t})}\left(\frac{\sigma}{2}\right)^{i-1} \bar{t}^{\frac{\nu}{2}-i+1}\sqrt{\pi}\ (-\zeta)^{i-1}J_{\nu-i+1}(\zeta),
\end{equation}
sharing precisely the same prefactors as the Bessel-$I$ functions. They can thus be taken out of the determinant and will cancel the prefactors in \eqref{detBs}. In the last step, when changing the index of the Bessel functions from decreasing to increasing, we obtain an extra sign in the first column originating from the $\hat{B}_i$, due to the identity \cite{NIST}
\begin{equation}
\zeta J_{\nu+1}(\zeta)=-\zeta J_{\nu-1}(\zeta)-2\nu J_\nu(\zeta).
\label{Bessel-Jid}
\end{equation}
For later convenience we will keep the first column sign free and thus move the sign into the columns with Bessel-$I$ functions, in both numerator and denominator.
Putting all results from this subsection together we arrive at the following limit for the unquenched kernel \eqref{kernel-final}
\begin{eqnarray}
\mathcal{K}_{A}^{(N_f)}(\zeta,\eta)&=&\lim_{N\rightarrow \infty} \frac{2\sqrt{|\zeta\eta|}}{4N \Xi} K_N^{(N_f)}\left(x=\frac{\zeta^2}{4N\Xi}, y = \frac{\eta^2}{4N \Xi} \right) 
\nonumber\\
&=&  \frac{\det \left[ \begin{matrix}
\mathcal{B}_{\text{JJ}}(\zeta,\eta) & \mathcal{B}_{\text{IJ}}(\mu_1,\eta) & \ldots &\mathcal{B}_{\text{IJ}}(\mu_{N_f},\eta) \\
J_\nu(\zeta) & I_\nu(\mu_1) & \ldots & I_\nu(\mu_{N_f}) \\
\zeta J_{\nu+1}(\zeta) & -\mu_1 I_{\nu+1}(\mu_1) & \ldots & -\mu_{N_f}I_{\nu+1}(\mu_{N_f}) \\
\vdots & \vdots & \ldots & \vdots \\
\zeta^{N_f-1}J_{\nu+N_f-1}(\zeta) & (-\mu_1)^{N_f-1}I_{\nu+N_f-1}(\mu_1) & \ldots & (-\mu_{N_f})^{N_f-1}I_{\nu+N_f-1}(\mu_{N_f}) \\
\end{matrix} \right]}{\sqrt{|\zeta\eta|}^{-1}\det\left[ (-\mu_f)^{j-1} I_{\nu+j-1}(\mu_f) \right]_{j,f=1}^{N_f}}, 
\nonumber\\
\label{final-kernelNf}
\end{eqnarray}
after dropping the prefactors in \eqref{kernel-final} that lead to an equivalent kernel, and multiplying the same Jacobian as in \eqref{finalkernel0}.
To summarise we have shown that the limiting unquenched kernel $\mathcal{K}_{A}^{(N_f)}(\zeta,\eta)$ of 
our polynomial ensemble can be made independent from $A$, as long as $t_c>1$ or equivalently the temperature dependent chiral condensate is non-vanishing, $\Xi>0$. It is thus universal. It remains to compare our new result to the representation of the $k$-point correlation functions from \cite{GuhrWettig} at $\nu=0$, which is not 
given by the determinant of a kernel. 
It is given by the ratio of determinants of sizes $N_f+k$ over $N_f$, respectively, and it is also independent of $A$ under the same conditions and thus also universal, when rescaling arguments and masses with $\Xi>0$. At $N_f=0$ the two results agree as was mentioned earlier. 
Furthermore, we will compare to the result \cite{DN,WGW} at $A=0$ which was derived at $\nu=0$, including a non-Gaussian weight function \cite{DN}. While their result is given by the determinant of a kernel, 
the limiting kernel is given by a ratio determinants of sizes $N_f+2$ over $N_f$, compared to $N_f+1$ over $N_f$ in our case. Again, in the quenched case $N_f=0$ the equivalence follows immediately from the Christoffel-Darboux identity. However, the inclusion of $N_f$ masses makes the comparison again a non-trivial task. Both comparisons are the subject of the next section.

\sect{Equivalence of zero and non-zero temperature results}
\label{sec:Universal}

We are now ready to compare our result for the unquenched limiting kernel at non-zero temperature to the known results in the literature. We will first show in the next subsection that \eqref{final-kernelNf}  agrees with the limiting kernel found in \cite{DN,WGW} at zero temperature, including $\nu \neq0$. In the subsequent subsection we will show that the $k$-point correlation function resulting from our kernel (and thus that of \cite{DN,WGW}) agree with those found in \cite{SWG} for non-zero temperature, using supersymmetry. Such an agreement was conjectured but so far only checked numerically in few cases. Both equivalences are non-trivial as they amount to compare representations in terms of determinants of different sizes. The key idea is to use a theorem from \cite{AV03} that yields different determinantal representations for the same expectation value of characteristic polynomials at finite-$N$, and in the second part, Subsection \ref{subsec:equivII}, to use a consistency condition proposed in \cite{ADII} in terms of limiting partition functions.

\subsection{Equivalence to the zero temperature kernel}
\label{subsec:equiv}

In this subsection we compare our result for the kernel \eqref{final-kernelNf} derived at non-zero temperature, which is  $A$-independent when measuring eigenvalues in units of $\Xi(A)>0$, to the universal kernel at zero temperature. 
We thus begin by quoting the known result \cite{DN,WGW}, the limiting kernel at zero temperature 
$A=0$, which we give here at $\nu\geq0$, cf. \cite{GA16}\footnote{There is a typo in \cite[Eq. (62)]{GA16} where the Jacobian from going to squared variables is missing.}:
\begin{eqnarray}
\mathcal{K}_0^{(N_f)}(\zeta,\eta)&=&\lim_{N\to\infty}\frac{2\sqrt{|\zeta\eta|}}{4N } \sqrt{w^{(N_f)}(x)w^{(N_f)}(y)}\widetilde{K}_{N,0}^{(N_f)}(x,y)\Big|_{x=\frac{\zeta^2}{4N}, y= \frac{\eta^2}{4N},m_f^2=\frac{\mu^2_f}{4N}} 
\nonumber\\
&=&
 \frac{\sqrt{\vert \zeta \eta \vert}}{\eta^2 -\zeta^2} \frac{\det \left[ 
\begin{matrix}
J_\nu(\zeta) & \zeta J_{\nu+1}(\zeta) & \ldots & \zeta^{N_f+1} J_{\nu+N_f+1}(\zeta) \\
J_\nu(\eta) & \eta J_{\nu+1}(\eta) & \ldots & \eta^{N_f+1} J_{\nu+N_f+1}(\eta) \\
I_\nu(\mu_1) & -\mu_1 I_{\nu+1}(\mu_1) & \ldots & (-\mu_1)^{N_f+1} I_{\nu+N_f+1}(\mu_1) \\
\vdots & \vdots & \ldots & \vdots \\
I_\nu(\mu_{N_f}) & -\mu_{N_f} I_{\nu+1}(\mu_{N_f}) & \ldots & (-\mu_{N_f})^{N_f+1} I_{\nu+N_f+1}(\mu_{N_f}) \\
\end{matrix}
\right]}{\prod_{f=1}^{N_f} \sqrt{(\zeta^2 + \mu_f^2)(\eta^2 + \mu_f^2)} \underset{1\leq f,g \leq N_f}{\det} \left[ (-\mu_f)^{g-1} I_{\nu+g-1}(\mu_f) \right]}.\quad\quad
\label{kernelDN}
\end{eqnarray}
It immediately leads to the microscopic density at $A=0$ when setting $\eta=\zeta$ and applying l'Hopital's rule:
\begin{eqnarray}
\mathcal{R}_{1,0}^{(N_f)}(\zeta)
&=&
- \frac{{\vert \zeta\vert}}{2} \frac{\det \left[ 
\begin{matrix}
\zeta^{-1}J_{\nu-1}(\zeta) & J_{\nu}(\zeta) & \ldots & \zeta^{N_f} J_{\nu+N_f}(\zeta) \\
J_\nu(\zeta) & \zeta J_{\nu+1}(\zeta) & \ldots & \zeta^{N_f+1} J_{\nu+N_f+1}(\zeta) \\
I_\nu(\mu_1) & -\mu_1 I_{\nu+1}(\mu_1) & \ldots & (-\mu_1)^{N_f+1} I_{\nu+N_f+1}(\mu_1) \\
\vdots & \vdots & \ldots & \vdots \\
I_\nu(\mu_{N_f}) & -\mu_{N_f} I_{\nu+1}(\mu_{N_f}) & \ldots & (-\mu_{N_f})^{N_f+1} I_{\nu+N_f+1}(\mu_{N_f}) \\
\end{matrix}
\right]}{\prod_{f=1}^{N_f} {(\zeta^2 + \mu_f^2)} \underset{1\leq f,g \leq N_f}{\det} \left[ (-\mu_f)^{g-1} I_{\nu+g-1}(\mu_f) \right]}.\quad\quad
\label{densityDN}
\end{eqnarray}
This limiting unquenched density agrees with \cite{DN} at $\nu=0$ and confirms the expression given in \cite{WGW} for $\nu>0$ without derivation. The fact that it is positive for all $\nu$ can be checked by taking successively the large mass limit, leading eventually to the quenched density \eqref{rhonu} which is known to be positive. 

Let us recall that at zero temperature we have set the parameter corresponding to the chiral condensate $\Sigma$ to unity. 
This result was derived in \cite{DN} at $\nu=0$ for a non-Gaussian potential and in \cite{WGW} for a Gaussian potential, where also the density was given for $\nu\neq0$. 
In the former, the factor $2\pi\rho(0)$ rescaling the eigenvalues was kept, to show the influence of the non-Gaussian potential through the mean density $\rho(0)$ at the origin, which is proportional to the chiral condensate $\Sigma$ through the Banks-Casher relation. 

The asymptotic result \eqref{kernelDN} follows straightforwardly from the finite-$N$ result we already quoted, when combining \eqref{kernel+w} and \eqref{Kernelmassive} and applying the standard asymptotic \cite[8.978.2]{Grad}
\begin{eqnarray}
\lim_{N\to\infty} N^{-\nu}L_n^{\nu}\left(\frac{\zeta^2}{4N}\right)&=&2^\nu \zeta^{-\nu}
J_\nu(\zeta),
\nonumber\\
\lim_{N\to\infty} N^{-\nu}L_n^{\nu}\left(\frac{-\mu^2}{4N}\right)&=&2^\nu \mu^{-\nu}
I_\nu(\zeta),
\label{Bessellim}
\end{eqnarray}
to the quenched orthogonal polynomials \eqref{Laguerre}. For completeness and later use we give the asymptotic of the polynomial part of the kernel  \eqref{KNsum}, and its analytic continuation in the first argument:
\begin{eqnarray}
\lim_{N\to\infty} N^{-1-\nu}\widetilde{K}_{N,0}^{(0)}\left(\frac{\zeta^2}{4N},\frac{\eta^2}{4N}\right)&=&2\frac{4^\nu}{(\zeta\eta)^\nu}\frac{\zeta J_{\nu+1}(\zeta)J_\nu(\eta)-\eta J_{\nu+1}(\eta)J_\nu(\zeta)}{\zeta^2-\eta^2},
\nonumber\\
\lim_{N\to\infty} N^{-1-\nu}\widetilde{K}_{N,0}^{(0)}\left(\frac{-\mu^2}{4N},\frac{\eta^2}{4N}\right)&=&2\frac{4^\nu}{(\mu\eta)^\nu}\frac{\mu I_{\nu+1}(\mu)J_\nu(\eta)-\eta J_{\nu+1}(\eta)I_\nu(\mu)}{\mu^2+\eta^2}.
\label{kernellim}
\end{eqnarray}
Recall that in order to obtain the limiting Bessel-kernel, the polynomial part of the kernel has to be multiplied by the weight function, see \eqref{kernel+w}. This cancels the $\nu$-dependent prefactors.

The equivalence can be shown by first deriving a different determinantal expression for the kernel at finite-$N$ that follows from \cite{AV03}, and then taking the large-$N$ limit that leads to \eqref{final-kernelNf}. We start with the expression of the polynomial part of the kernel in terms of expectation values of characteristic polynomials \eqref{KNdet2} at finite-$N$,
that we repeat here for convenience:
\begin{equation}
\label{KNvev}
\widetilde{K}_{N,0}^{(N_f)}(x,y)
=\frac{(-1)^{N_f}}{h_{N-1}^{(0)}}\frac{\mathbb{E}_{\mathcal{P}_{N-1,\nu}^{0}}\left[D_{N-1}(x)D_{N-1}(y)\prod_{f=1}^{N_f}D_{N-1}(-m_f^2)\right] }{\mathbb{E}_{\mathcal{P}_{N,\nu}^{0}}\left[\prod_{f=1}^{N_f}D_{N}(-m_f^2)\right]}.
\end{equation}
Notice the different dimensions $N-1$ respectively $N$ in the expectation values.
In \cite{AV03} the following theorem was proven for expectation values of products of characteristic polynomials, which we give here for Hermitian random matrix ensembles, in terms 
of
monic (quenched) polynomials $p_k^{(0)}(x)$ instead of general orthonormal polynomials as in \cite{AV03}:
\begin{eqnarray}
\mathbb{E}_{\mathcal{P}_{N,\nu}^{0}}\left[\prod_{k=1}^KD_{N}(v_k)\prod_{l=1}^{L}D_{N}(u_l)\right]&=& \frac{\prod_{j=N}^{N+L-1}h_j^{(0)}}{\Delta_{K}(\{v\})\Delta_{L}(\{u\})}
\nonumber\\
&&
\times\det\left[
\begin{matrix}
\widetilde{K}_{N+L,0}^{(0)}(v_1,u_1)&\ldots&\widetilde{K}_{N+L,0}^{(0)}(v_K,u_1)\\
\vdots&\ldots&\vdots\\
\widetilde{K}_{N+L,0}^{(0)}(v_1,u_L)&\ldots&\widetilde{K}_{N+L,0}^{(0)}(v_K,u_L)\\
p_{N+L}^{(0)}(v_1)& \ldots& p_{N+L}^{(0)}(v_K)\\
\vdots&\ldots&\vdots\\ 
p_{N+K-1}^{(0)}(v_1)& \ldots& p_{N+K-1}^{(0)}(v_K)\\
\end{matrix}
\right],
\label{ThmAV}
\end{eqnarray}
for $K\geq L$, without loss of generality. We also have transposed the matrix inside the determinant, for later convenience. Initially, this identity was derived in non-Hermitian ensembles, hence the split into a product of determinants and a product of conjugated determinants. As it was remarked already in \cite{AV03}, when applying it to Hermitian ensembles there are many possible ways to split a product of $K+L$ characteristic polynomials into two groups. This gives rise to an entire family of equivalent expressions in terms of determinants of different sizes, with a different number of kernels and polynomials\footnote{The reason why this is not possible in non-Hermitian ensembles is the lack of an appropriate Christoffel-Darboux formula for the kernel of planar orthogonal polynomials.}. 

Coming back to our problem, in the derivation of \eqref{Kernelmassive} the choice $K=N_f+2$ and $L=0$ was made in the numerator (and $K=N_f$ and $L=0$ in the denominator). Keeping the same choice for the denominator, we will now choose $K=N_f+1$ and $L=1$ in the numerator, with $v_1,\ldots,v_K=-m_1^2,\ldots,-m_{N_f}^2$, and $v_{K+1}=x$ as well as $u_1=y$. This immediately leads to the following, alternative representation of the polynomial part \eqref{KNsum} of the zero temperature kernel compared to \eqref{Kernelmassive}:
\begin{equation}
\widetilde{K}_{N,0}^{(N_f)}(x,y)=(-1)^{N_f}\frac{
\det\left[
\begin{matrix}
\widetilde{K}_{N,0}^{(0)}(-m_1^2,y)&\ldots&\widetilde{K}_{N,0}^{(0)}(-m_{N_f}^2,y)
&\widetilde{K}_{N+L,0}^{(0)}(x,y)\\
L_{N}^{\nu}(-m_1^2)& \ldots& L_{N}^{\nu}(-m_{N_f}^2)&L_{N}^{\nu}(x)\\
\vdots&\ldots&\vdots&\vdots\\ 
L_{N+N_f-1}^{\nu}(-m_1^2)& \ldots& L_{N+N_f-1}^{\nu}(-m_{N_f}^2)&L_{N+N_f-1}^{\nu}(x)\\
\end{matrix}
\right]
}{\prod_{f=1}^{N_f}(x+m_f^2)
\det[L_{N+j-1}^\nu(-m_i^2)]_{i,j=1}^{N_f}}.
\label{Kernelmassive2}
\end{equation}
The kernel in the first line on the right hand side is the quenched kernel in the Christoffel-Darboux form in terms of Laguerre polynomials, \eqref{KNsum} at $N_f=0$. It is not difficult now to take the large-$N$ limit of \eqref{Kernelmassive2}. Upon choosing the an appropriate equivalence factor for the kernel,  we obtain the following limit: 
\begin{eqnarray}
\mathcal{K}_0^{(N_f)}(\zeta,\eta)&=&
\lim_{N\to\infty}\frac{2\sqrt{|\zeta\eta|}}{4N} \sqrt{w^{(N_f)}(x)w^{(N_f)}(y)}
\sqrt{\prod_{f=1}^{N_f}\frac{y+m_f^2}{x+m_f^2}}\ 
\widetilde{K}_{N,0}^{(N_f)}(x,y)\Big|_{x=\frac{\zeta^2}{4N}, y= \frac{\eta^2}{4N},m_f^2=\frac{\mu^2_f}{4N}} 
\nonumber\\
&=&
\mathcal{K}_{A}^{(N_f)}(\zeta,\eta).
\end{eqnarray}
Here, we moved the last column in \eqref{Kernelmassive2} to become the first, and applied the asymptotic \eqref{Bessellim} and \eqref{kernellim} to the polynomials and kernels. After cancelling factors from the determinant in the denominator and numerator we arrive at \eqref{final-kernelNf}. This is the equivalence we wanted to prove, showing that the limiting kernel at non-zero temperature \eqref{final-kernelNf} is equivalent to the universal kernel \eqref{kernelDN} at zero temperature, and that this equivalence extends from previous results to non-zero topology $\nu>0$ as well. We mention in passing that further equivalent forms exist for the same kernel, at finite and infinite $N$, depending on how the product of characteristic polynomials in the numerator of \eqref{KNvev} is split into two.

\subsection{Equivalence with the $k$-point correlation functions at non-zero temperature}
\label{subsec:equivII}

In this subsection we compare with the result for the unquenched $k$-point correlation functions at non-zero temperature \cite{SWG}. They are given in terms of a determinant of size $N_f+k$, and below we will prove that it agrees with the determinantal expression of size $k$ in terms of the unquenched kernel at zero temperature (which is itself a determinant). In fact we will state the expression for the $k$-point function for $\nu\geq0$, generalising the result determined in \cite{SWG} for $\nu=0$. Taking the microscopic limit of the $k$-point functions defined in \eqref{RNkdef} at the origin, it holds  
\begin{eqnarray}
&&\rho_{k,A}^{(N_f)}(\zeta_1,\ldots,\zeta_k)\equiv \lim_{N\to\infty}
\frac{2^k\prod_{j=1}^k|\zeta_j|}{(4N\Xi)^k}R_{k,A}^{(N_f)}(x_1,\ldots,x_k)\Big|_{x_k=\frac{\zeta_k^2}{4N\Xi}}\label{rhokdef}\\
&&=\prod_{j=1}^k|\zeta_j|
\frac{\det \left[ \begin{matrix}
\mathcal{B}_{\text{JJ}}(\zeta_1,\zeta_1) &\ldots & \mathcal{B}_{\text{JJ}}(\zeta_1,\zeta_k) &J_\nu(\zeta_1) & 
\ldots &\zeta_1^{N_f-1}J_{\nu+N_f-1}(\zeta_1)\\
\vdots &\ldots & \vdots & 
\ldots &\vdots \\
\mathcal{B}_{\text{JJ}}(\zeta_k,\zeta_1) &\ldots & \mathcal{B}_{\text{JJ}}(\zeta_k,\zeta_k) &J_\nu(\zeta_k) & 
\ldots & \zeta_k^{N_f-1}J_{\nu+N_f-1}(\zeta_k)\\
\mathcal{B}_{\text{IJ}}(\mu_1,\zeta_1)  &\ldots &\mathcal{B}_{\text{IJ}}(\mu_1,\zeta_k) &  I_\nu(\mu_1) &
\ldots &(-\mu_1)^{N_f-1}I_{\nu+N_f-1}(\mu_1)\\
\vdots &\ldots & \vdots & 
\ldots &\vdots \\
\mathcal{B}_{\text{IJ}}(\mu_{N_f},\zeta_1) &\ldots &\mathcal{B}_{\text{IJ}}(\mu_{N_f},\zeta_k) &  I_\nu(\mu_{N_f}) & 
\ldots & 
 (-\mu_{N_f})^{N_f-1}I_{\nu+N_f-1}(\mu_{N_f}) \\
\end{matrix} \right]}{\det\left[ (-\mu_f)^{j-1} I_{\nu+j-1}(\mu_f) \right]_{j,f=1}^{N_f}}, \quad\quad
\label{rhokSGW}
\end{eqnarray}
where $\mathcal{B}_{IJ}$ and $\mathcal{B}_{JJ}$ are defined in \eqref{BIJdef} and \eqref{KBdef}, respectively.

We will pursue the following strategy:
It is known \cite{ADII} that for orthogonal polynomials ensembles, that is at $A=0$, the limiting $k$-point correlation function can be written not only as the $k\times k$ determinant of the limiting kernel, but also in terms of the limiting partition function with $2k$ additional flavours
\begin{equation}
\rho_{k,0}^{(N_f)}(\zeta_1,\ldots,\zeta_k)= (-1)^{k\nu}\prod_{j=1}^k\!\left(\!|\zeta_j|\prod_{f=1}^{N_f}(\zeta_j^2+\mu_f^2)\!\right)\!\Delta_k(\{\zeta^2\})^2
\frac{\mathcal{Z}_\nu^{(N_f+2k)}(\mu_1\ldots,\mu_{N_f},i\zeta_1,i\zeta_1,\ldots,i\zeta_k,i\zeta_k)}{\mathcal{Z}_{\nu}^{(N_f)}(\mu_1\ldots,\mu_{N_f})}.
\label{rhokZNf2k}
\end{equation}
We explicitly spelled out the arguments of the partition functions on the right hand side, to indicate the twofold degeneracy of the $2k$ flavours. 
Regarding the sign, a partition function of imaginary masses is not necessarily positive. 
In fact the proportionality constant $(-1)^{k\nu}$ follows from the required positivity of the $k$-point function, that can be traced back to \eqref{densityDN} as we will show below. This constant was left undetermined in \cite{ADII}. 
We note in passing that in view of \eqref{ZNvev} this yields yet another representation as a determinant of size $N_f+2k$ \cite{ADII}, compared to size $N_f+k$ in \eqref{rhokSGW}, and we will also show these to be equivalent. 

Consequently, we have to determine the asymptotic of the partition function $\mathcal{Z}_{\nu}^{(N_f)}(\{\mu\})$ as it appears in the statement \eqref{rhokZNf2k}
\begin{equation} 
\lim_{N\to\infty}Z_{N,\nu}^{(N_f)}(\{m\})\Big|_{m_f^2=\frac{\mu_f^2}{4N}}\sim 
\frac{\det\left[ \mu_f^{j-1} I_{\nu+j-1}(\mu_f) \right]_{j,f=1}^{N_f}}{\Delta_{N_f}(\{\mu^2\})}\equiv \mathcal{Z}_{\nu}^{(N_f)}(\{\mu\}).
\label{Zlimdef}
\end{equation} 
Here and in the following we will suppress the $N$-dependent factor needed to arrive at the right hand side. For example, from \eqref{ZNvev} at $N_f=1$, we would have $Z_{N,\nu}^{(N_f=1)}(m)/Z_{N,\nu}^{(0)}\sim N!N^{\nu/2} I_\nu(\mu)$. We are only interested in the finite part that depends on the rescaled masses. 
The normalisation in \eqref{Zlimdef} is chosen such that it 
is positive and 
agrees with the finite volume partition function at fixed topology from the $\varepsilon\chi PT$ regime 
\cite{JSVgroup,JSVgroup2}
\begin{equation}
\int_{U(N_f)}d[U]\det[U]^\nu\exp\left[\frac12 V\Sigma\tr(M(U+U^\dag))\right]=\frac{\det\left[ \mu_f^{j-1} I_{\nu+j-1}(\mu_f) \right]_{j,f=1}^{N_f}}{\Delta_{N_f}(\{\mu^2\})},
\end{equation}
when identifying $m_fV\Sigma=\mu_f$ in the diagonal mass matrix of the quarks $M=\diag(m_1,\ldots,m_{N_f})$. The integration $d[U]$ is over the appropriately normalised Haar measure of the unitary group.

In a first step we will show, that \eqref{rhokSGW} agrees with \eqref{rhokZNf2k} in terms of the finite volume partition functions.
Then, we can insert  the following consistency condition proposed in \cite{ADII} (up to a proportionality constant) into \eqref{rhokZNf2k}: 
\begin{equation}
\Delta_k(\{\xi^2\})\Delta_k(\{\eta^2\})\frac{\mathcal{Z}_\nu^{(N_f+2k)}(\{\mu\},\xi_1,\ldots,\xi_k,\eta_1,\ldots,\eta_k)}{\mathcal{Z}_{\nu}^{(N_f)}(\{\mu\})}=\det\left[\frac{\mathcal{Z}_\nu^{(N_f+2)}(\{\mu\},\xi_a,\eta_b)}{\mathcal{Z}_{\nu}^{(N_f)}(\{\mu\})}\right]_{a,b=1}^k.
\label{CCI}
\end{equation}
It relates finite volume partition functions of the two sets of variables $\{\xi_1,\ldots,\xi_k\}$ and $\{\eta_1,\ldots,\eta\}$, and was later proven in \cite{Braden} based on Wick's theorem. 
Setting the two sets of variables equal to $\{i\zeta_1,\ldots,i\zeta_k\}$, we can then identify the kernel as the limiting partition function with two additional flavours $N_f+2$ under the determinant on the right hand side \cite{ADII}, 
which essentially follows from combining \eqref{KNreweight} and \eqref{ZNvev}. This leads to the determinantal expression of the $k$-point function using the kernel \eqref{kernelDN} which is equivalent to our result \eqref{final-kernelNf}, as we have already shown in the previous subsection. This establishes then the equivalence between our result based on polynomial ensembles and \cite{SWG} based on supersymmetry, including its extension to $\nu\geq0$.

It is quite possible that the identity \eqref{CCI} can be shown to hold for expectation values of ratios of characteristic polynomials at finite-$N$. And perhaps this remains even true in polynomial ensembles, see e.g. \cite[Theorem 2.3]{ASW} for ratios of an equal number of characteristic polynomials in such a setting.

Let us begin by computing the limiting partition function on the right hand side of \eqref{rhokZNf2k}, where we use the first line of \eqref{ZNvev} as a starting point,
\begin{eqnarray}
\label{ZN2kexp}
&&\frac{Z_{N,\nu}^{(N_f+2k)}(\{m\},\{iz,iz\})}{Z_{N,\nu}^{(0)}} =  (-1)^{N(N_f+2k)+k\nu} 
\prod_{f=1}^{N_f}m_f^\nu\prod_{j=1}^kz_j^{2\nu}\ 
\mathbb{E}_{\mathcal{P}_{N,\nu}^0}\! \!\left[ \prod_{f=1}^{N_f} D_N(-m_f^2) \prod_{m=1}^{k} D_N(z_m^2)^2\right]\nonumber\\
&=&\frac{(-1)^{NN_f+k\nu} \prod_{f=1}^{N_f}m_f^\nu\prod_{j=1}^kz_j^{2\nu}\prod_{j=N}^{N+k-1}h_j^{(0)}}{\left(\prod_{j=1}^k\prod_{f=1}^{N_f}(z_j^2+m_f^2)\right)\Delta_{N_f}(\{-m^2\})\Delta_k(\{z^2\})^2}
\nonumber\\
&&\times\det\left[
\begin{matrix}
\widetilde{K}_{N+k,0}^{(0)}(-m_1^2,z_1^2)&\ldots&\widetilde{K}_{N+k,0}^{(0)}(-m_{N_f}^2,z_1^2)&\widetilde{K}_{N+k,0}^{(0)}(z_1^2,z_1^2)&\ldots&\widetilde{K}_{N+k,0}^{(0)}(z_k^2,z_1^2)\\
\vdots&\ldots&\vdots&\vdots&\ldots&\vdots\\
\widetilde{K}_{N+k,0}^{(0)}(-m_1^2,z_k^2)&\ldots&\widetilde{K}_{N+k,0}^{(0)}(-m_{N_f}^2,z_k^2)
&\widetilde{K}_{N+k,0}^{(0)}(z_1^2,z_k^2)&\ldots&\widetilde{K}_{N+k,0}^{(0)}(z_k^2,z_k^2)\\
p_{N+k}^{(0)}(-m_1^2)& \ldots& p_{N+k}^{(0)}(-m_{N_f}^2)
&p_{N+k}^{(0)}(z_1^2)& \ldots& p_{N+k}^{(0)}(z_k^2)\\
\vdots&\ldots&\vdots&\vdots&\ldots&\vdots\\ 
p_{N+k+N_f-1}^{(0)}(-m_1^2)& \ldots& p_{N+k+N_f-1}^{(0)}(-m_{N_f}^2)&p_{N+k+N_f-1}^{(0)}(z_1^2)& \ldots& p_{N+k+N_f-1}^{(0)}(z_k^2)\\
\end{matrix}
\right].\nonumber\\
\end{eqnarray}
In the second step  we have used \eqref{ThmAV} in choosing $K=N_f+k$ and $L=k$.
Next, we take the scaling limit of this expression divided by $Z_{N,\nu}^{(N_f)}(\{m\})$ from \eqref{ZNvev}.
Using the explicit form of the monic orthogonal polynomials  \eqref{Laguerre},  we can apply the asymptotic \eqref{Bessellim} and \eqref{kernellim} in terms of the rescaled variables $m_f^2=\mu_f^2/(4N)$ and $z_k^2=\zeta_k^2/(4N)$. Ignoring the prefactor in powers of $N$ we obtain 
\begin{eqnarray}
&&\frac{\mathcal{Z}_{\nu}^{(N_f+2k)}(\{m\},\{iz,iz\})}{\mathcal{Z}_{\nu}^{(N_f)}(\{m\})}
=\frac{(-1)^{N_fk+k\nu}}{\det\left[ (-\mu_f)^{j-1} I_{\nu+j-1}(\mu_f) \right]_{j,f=1}^{N_f}\left(\prod_{j=1}^k\prod_{f=1}^{N_f}(\zeta_j^2+\mu_f^2)\right)\Delta_k(\{\zeta^2\})^2}
\nonumber\\
&&\times\det\left[
\begin{matrix}
\mathcal{B}_{\text{IJ}}(\mu_1,\zeta_1)  &\ldots &\mathcal{B}_{\text{IJ}}(\mu_1,\zeta_k) &  I_\nu(\mu_1) &
\ldots &(-\mu_1)^{N_f-1}I_{\nu+N_f-1}(\mu_1)\\
\vdots &\ldots & \vdots & 
\ldots &\vdots \\
\mathcal{B}_{\text{IJ}}(\mu_{N_f},\zeta_1) &\ldots &\mathcal{B}_{\text{IJ}}(\mu_{N_f},\zeta_k) &  I_\nu(\mu_{N_f}) & 
\ldots & 
 (-\mu_{N_f})^{N_f-1}I_{\nu+N_f-1}(\mu_{N_f}) \\
 \mathcal{B}_{\text{JJ}}(\zeta_1,\zeta_1) &\ldots & \mathcal{B}_{\text{JJ}}(\zeta_1,\zeta_k) &J_\nu(\zeta_1) & 
\ldots &\zeta_1^{N_f-1}J_{\nu+N_f-1}(\zeta_1)\\
\vdots &\ldots & \vdots & 
\ldots &\vdots \\
\mathcal{B}_{\text{JJ}}(\zeta_k,\zeta_1) &\ldots & \mathcal{B}_{\text{JJ}}(\zeta_k,\zeta_k) &J_\nu(\zeta_k) & 
\ldots & \zeta_k^{N_f-1}J_{\nu+N_f-1}(\zeta_k)\\
\end{matrix}
\right].\nonumber\\
\end{eqnarray}
In the last line we transposed the matrix under the determinant.
This gives an alternative representation of the partition function with $N_f+2k$ flavours compared to \cite{ADII}, which results from the direct limit of \eqref{ZNvev} (with $K=N_f+2k$ and $L=0$).
 Permuting the $\zeta_j$-dependent rows to become the first rows, we can insert this expression into \eqref{rhokZNf2k} and observe that almost all prefactors of the determinant cancel. Consequently, we have just derived that \eqref{rhokSGW} and \eqref{rhokZNf2k} agree,
\begin{equation}
\rho_{k,A}^{(N_f)}(\zeta_1,\ldots,\zeta_k)=\rho_{k,0}^{(N_f)}(\zeta_1,\ldots,\zeta_k),
\end{equation}
that is the non-zero temperature $k$-point function from \cite{SWG} and the zero temperature one resulting from \eqref{rhokZNf2k} agree.

As a last step we insert the consistency condition \eqref{CCI} at $\xi_j=\eta_j=i\zeta_j$ for all $j=1,\ldots,k$ into \eqref{rhokZNf2k}, to give
\begin{eqnarray}
\rho_{k,0}^{(N_f)}(\zeta_1,\ldots,\zeta_k)&=& (-1)^{k\nu}\prod_{j=1}^k\left(|\zeta_j|\prod_{f=1}^{N_f}(\zeta_j^2+\mu_f^2)\right)
\det\left[\frac{\mathcal{Z}_\nu^{(N_f+2)}(\{\mu\},i\zeta_a,i\zeta_b)}{\mathcal{Z}_{\nu}^{(N_f)}(\{\mu\})}\right]_{a,b=1}^k
\nonumber\\
&=& \det\left[\mathcal{K}_{0}^{(N_f)}(\zeta_a,\zeta_b)\right]_{a,b=1}^k.
\label{finalequivII}
\end{eqnarray}
In the second equation we have inserted the expression for the unquenched limiting kernel \eqref{kernelDN} in terms of the ratio of finite volume partition functions with two extra flavours from \cite{ADII},
\begin{equation}
\label{Krelation}
\mathcal{K}_{0}^{(N_f)}(\zeta,\eta)=(-1)^{\nu}\sqrt{|\zeta\eta|}\prod_{f=1}^{N_f}\sqrt{(\zeta^2+\mu_f^2)(\eta^2+\mu_f^2)} \frac{\mathcal{Z}_\nu^{(N_f+2)}(\{\mu\},i\zeta,i\eta)}{\mathcal{Z}_{\nu}^{(N_f)}(\{\mu\})}.
\end{equation}
It follows combining \eqref{KNreweight} and \eqref{ZNvev} and in consequence in \eqref{finalequivII} all factors cancel out. Our result \eqref{Krelation} differs from 
\cite{ADII} by a factor of $(-1)^{[N_f/2]+1}$ on the right hand side, after taking into account the different convention for the Vandermonde determinant in \cite{ADII}. 
The sign therein thus seems to be incorrect, as we obtain a positive density \eqref{densityDN} from \eqref{Krelation} for all values of $\nu$ and $N_f$.
In summary we have established that the result \eqref{rhokSGW} can be also written in terms of a determinant of the limiting kernel(s) of size $k\times k$, and that thus the two results from non-zero temperature are equivalent as well. 


\sect{Summary and outlook}\label{sec:conclusio}

In this work we have revisited the random matrix ensemble for the low energy Dirac operator spectrum at non-zero temperature, including an arbitrary  fixed number of quark flavours at non-zero topology. 
We have established that its eigenvalue correlation functions can be written as the determinant of a kernel, being part of polynomial ensembles in random matrix theory. 
Using very recent results of the authors, the corresponding kernel was constructed in terms of a double contour integral. Its 
microscopic origin 
limit was evaluated through a saddle point analysis, where we could  follow similar strategies from  previous works. 
When rescaling the eigenvalues with the temperature dependent chiral condensate below the critical temperature, we were able to establish a universal answer for the limiting kernel and resulting $k$-point correlations functions. It depends only on the quark masses and topology, but not on the parameters modelling the effect of temperature. 

In the second part we made contact to previous universality results, first to the zero temperature case, where the link between random matrix theory and the leading order of the $\varepsilon$-regime in chiral perturbation theory has been first established. Our kernel was shown to be equivalent to the known universal kernel at zero temperature, although being given by a ratio of determinants of different sizes. 
Furthermore, we could establish that the $k$-point correlation functions, that were computed previously in the ensemble with temperature using supersymmetric techniques, agree with the determinantal expression of the kernel with (and without) temperature, that follows from the polynomial ensemble.
The tools that we used  for proving these equivalences was, first, a theorem that expresses the expectation value of the product of characteristic polynomials in  orthogonal polynomial ensembles in various equivalent forms of determinants of different sizes, containing the quenched kernel and orthogonal polynomials of the underlying ensemble only. Second, we used a set of consistency condition for finite volume partition functions that also relates determinantal expressions of different sizes. In passing, we showed that also these partition functions can take different determinantal representations.

Several open questions seem to be interesting to pursue: First, it is tempting to tune the parameters representing temperature to the chiral phase transition, in order to make a detailed analysis of the microscopic kernel and correlations functions of the QCD Dirac operator at criticality. The framework of a polynomial ensemble and our detailed knowledge of the kernel at finite $N$, including its dependence on topology, make this a promising task. To our knowledge this has not been done beyond the quenched microscopic density yet. 

The analysis of the random matrix ensemble we have studied was recently used to deduce information about non-trivial eigenvector correlations in non-Hermitian ensembles, in particular for the real eigenvalues of the real Ginibre ensemble that appears in many applications. It will be very interesting to see if our results can be further applied in this direction. 

A more mathematical question concerns possible identities among expectation values of products or ratios of characteristic polynomials in polynomial ensembles, for finite or infinite matrix size. In our approach, we have been obliged to first establish equivalence with the ensemble at zero temperature, where many relations amongst these expectation values are known. It would be very interesting to know if such identities exist directly on the level of polynomial ensembles. First steps in this direction in our previous work are very encouraging.

\section*{Acknowledgments}

We would like to thank Eugene Strahov for useful discussions and correspondence. 
This work was funded by the Deutsche Forschungsgemeinschaft (DFG, German Research Foundation) – SFB 1283/2 2021 – 317210226 "Taming uncertainty and profiting from randomness and low regularity in analysis, stochastics and their applications" (GA and TW) and  by EPSRC Grant
EP/V002473/1 (TW).
We thank the Department of Mathematics at KTH Stockholm for hospitality 
and partial funding through The Knut
and Alice Wallenberg Foundation (GA), 
where part of this work was developed. 
Last but not least we acknowledge partial support by an NSF grant No. DMS-1928930, while the one of authors (GA) participated in a program hosted
by the Mathematical Sciences Research Institute 
"Universality and Integrability in Random Matrix Theory and Interacting Particle Systems" during the Fall semester 2021. We thank Thomas Guhr for a critical reading of the manuscript.


\end{document}